\documentclass[10pt,journal,compsoc]{IEEEtran}
\usepackage{amsmath,amssymb,amsfonts}
\usepackage{bm}
\usepackage[nocompress]{cite}
\usepackage{graphicx}
\usepackage{amsmath}
\usepackage{algorithm}
\usepackage{algpseudocode} 
\usepackage{array} 
\usepackage{enumitem}
\usepackage[caption=false,font=footnotesize,labelfont=sf,textfont=sf]{subfig}
\usepackage{textcomp}
\usepackage{stfloats}
\usepackage{url}
\usepackage{verbatim}
\usepackage[dvipsnames]{xcolor}

\begin{document}
\title{Frame Structure and Protocol Design for Sensing-Assisted NR-V2X Communications}

\author{Yunxin~Li,~\IEEEmembership{Graduate~Student~Member,~IEEE,}
        Fan~Liu,~\IEEEmembership{Senior~Member,~IEEE,}
        Zhen~Du,~\IEEEmembership{Member,~IEEE,}
        Weijie~Yuan,~\IEEEmembership{Member,~IEEE,}
        Qingjiang~Shi,~\IEEEmembership{Senior~Member,~IEEE,}
        and~Christos~Masouros,~\IEEEmembership{Senior~Member,~IEEE}
\IEEEcompsocitemizethanks{
\IEEEcompsocthanksitem A part of this article was presented at the IEEE International Conference on Communications (ICC), Italy, May 2023 \cite{li2023isacenabled}.
\IEEEcompsocthanksitem Yunxin Li, Fan Liu and Weijie Yuan are with the Department of Electronic and Electrical Engineering, Southern University of Science and Technology, Shenzhen 518055, China. Fan Liu is also with Peng Cheng Laboratory, Shenzhen 518066, China.\protect\\
E-mail: liyx2022@mail.sustech.edu.cn, \{liuf6, yuanwj\}@sustech.edu.cn.
\IEEEcompsocthanksitem Zhen Du is with the School of Electronic and Information Engineering, Nanjing University of Information Science and Technology, Nanjing 210044, China, and is also with the Department of Electronic and Electrical Engineering, Southern University of Science and Technology, Shenzhen 518055, China. \protect\\
E-mail: duzhen@nuist.edu.cn.
\IEEEcompsocthanksitem Qingjiang Shi is with the School of Software Engineering, Tongji University, Shanghai 201804, China, and also with the Shenzhen Research Institute of Big Data, Shenzhen 518172, China. \protect\\
E-mail: shiqj@tongji.edu.cn.
\IEEEcompsocthanksitem Christos Masouros is with the Department of Electronic and Electrical Engineering, University College London, London WC1E 7JE, U.K. \protect\\
E-mail: c.masouros@ucl.ac.uk.
}
\thanks{(Corresponding author: Fan Liu.)}
}

\IEEEtitleabstractindextext{%
\begin{abstract}
The emergence of the fifth-generation (5G) New Radio (NR) technology has provided unprecedented opportunities for vehicle-to-everything (V2X) networks, enabling enhanced quality of services. However, high-mobility V2X networks require frequent handovers and acquiring accurate channel state information (CSI) necessitates the utilization of pilot signals, leading to increased overhead and reduced communication throughput. To address this challenge, integrated sensing and communications (ISAC) techniques have been employed at the base station (gNB) within vehicle-to-infrastructure (V2I) networks, aiming to minimize overhead and improve spectral efficiency. In this study, we propose novel frame structures that incorporate ISAC signals for three crucial stages in the NR-V2X system: initial access, connected mode, and beam failure and recovery. These new frame structures employ 75\% fewer pilots and reduce reference signals by 43.24\%, capitalizing on the sensing capability of ISAC signals. Through extensive link-level simulations, we demonstrate that our proposed approach enables faster beam establishment during initial access, higher throughput and more precise beam tracking in connected mode with reduced overhead, and expedited detection and recovery from beam failures. Furthermore, the numerical results obtained from our simulations showcase enhanced spectrum efficiency, improved communication performance and minimal overhead, validating the effectiveness of the proposed ISAC-based techniques in NR V2I networks.
\end{abstract}

}

\maketitle
\IEEEdisplaynontitleabstractindextext
\IEEEpeerreviewmaketitle
\ifCLASSOPTIONcompsoc
\IEEEraisesectionheading{\section{Introduction}\label{sec:introduction}}
\else
\section{Introduction}
\label{sec:introduction}
\fi

\IEEEPARstart{V}{2X}  networks have emerged as a transformative technology with the potential to revolutionize transportation systems by enabling vehicles to establish communications, not only amongst themselves but also with surrounding infrastructure, pedestrians and other road users. At the core of this paradigm shift lies a robust communication framework that facilitates seamless and efficient data exchange, thereby enhancing road safety, optimizing traffic flow and improving overall mobility experiences. To enable effective communication modes, V2X networks rely on a combination of wireless technologies. Two prominent technologies deployed in V2X networks are Dedicated Short-Range Communications (DSRC) \cite{abboud2016interworking} \cite{ma2020joint} and Cellular V2X (C-V2X) \cite{garcia2021tutorial}. However, despite their advantages, both technologies come with inherent limitations. DSRC, for instance, is restricted to a dedicated frequency band (5.9 \text{GHz}), which poses scalibility challenges due to limited bandwidth. Additionally, sharing this frequency band with other wireless services introduces potential interference and coexistence issues. Although the 3rd
Generation Partnership Project (3GPP) published specifications on NR sidelink, focusing on improving the latency, capacity and flexibility of V2X since Release 16 \cite{3gpp.22.186}\cite{3gpp.38.885}, the reliability and accuracy of NR-V2X sidelink positioning still requires hybrid approaches such as global navigation satellite-based systems (GNSS) to achieve more robust and precise location estimation, which suffers from limited refresh rate. Moreover, the high mobility inherent in V2X networks presents significant challenges in channel training and beamforming, necessitating frequent coordination and feedback between the gNB and the vehicles. This poses critical issues in terms of signaling overhead and may lead to a degradation in communication throughput. 

Among the emerging technologies, millimeter-wave (mmWave) and massive multiple-input-multiple-output (mMIMO) present promising opportunities to overcome the limitations of traditional V2X communication with significantly improved performance \cite{va2016millimeter}\cite{heath2016overview}. The increased bandwidth in mmWave operating above 30 GHz enables the transmission of larger amounts of data as well as higher resolution for sensing. In conjunction with mmWave, mMIMO array equipped with a large number of antennas can form highly directional beams and create spatially separated communication channels, enabling simultaneous data transmission to multiple V2X nodes. More relavant to this work, the concept of integrated sensing and communications (ISAC) has garnered attention as a promising solution that fully reaps the benefits of mmWave and mMIMO technologies in NR V2X networks \cite{liu2020joint}. Recent studies have reported significant advantages of ISAC signaling in reducing channel estimation overhead compared to conventional mmWave beam training and tracking techniques used solely for communication purposes\cite{liu2020radar}\cite{du2022integrated}\cite{gonzalez2016radar}. This is primarily due to the removal of the necessity for dedicated downlink pilots in ISAC signaling, as well as eliminating the need for uplink feedback and the resulting transmission and quantization errors, through direct processing of the reflected echoes from the vehicles. These findings highlight the promising prospects of ISAC signaling in optimizing the overall performance of mmWave-based V2X systems.

Recent research efforts in ISAC V2X networks have focused on designing advanced techniques and algorithms in beam training and tracking to enhance the performance of V2X systems. These efforts aim to leverage the sensing capabilities for serving communication links, thereby improving the reliability and efficiency of vehicular networks. Predictive beamforming approaches based on extended Kalman filtering (EKF) \cite{liu2020radar}, Bayesian message passing algorithm \cite{yuan2020bayesian} and deep learning \cite{mu2021integrated} were proposed to increase the precision of beam alignment while reducing the beam training overhead associated with pilots. To address the challenges posed by extended vehicular targets, a dynamic predictive beamforming technique, namely, ISAC-AB, was proposed in \cite{du2022integrated}, which adaptively adjusts the beamwidth based on the location of the targets. Furthermore, roadway-geometry aware ISAC beam tracking method was designed to remain connectivity on complex trajectories \cite{meng2022vehicular}. These ISAC-based beam tracking and beamforming algorithms offer multiple advantages compared to conventional methods based on communication-only specifications that require pilot signals and feedback \cite{lim2020efficient}\cite{tan2021wideband}. During tracking, ISAC system employs echo signals to analyze the parameters of interest pertaining to the vehicles, as opposed to relying on periodic channel state information (CSI) feedback from the vehicles. This approach ensures uninterrupted downlink data transmission and enables precise measurement of the targets, consequently mitigating the overhead associated with pilot signals and feedback. Additionally, the utilization of all downlink frames for both sensing and communications enhances the matched-filtering gain, thereby improving the accuracy of sensing, the resulting beamforming prediction and ensuring reliable communication quality. The sensing capability of the ISAC system extends beyond the angular domain, encompassing localization of the vehicles in terms of range and speed. This comprehensive sensing ability enables more precise localization of the vehicles. 

Despite the considerable body of research on ISAC-enabled V2X networks, the application of these approaches within the framework of 5G NR presents challenges due to the complex frame structures and transmission protocols inherent in 5G systems. Previous studies often assume simplified frame structures for ease of analysis, rendering their direct applicability to the intricate 5G NR framework uncertain. More importantly, while theoretical studies demonstrate the potential for reducing overhead through the utilization of echo signals and by leveraging the inherent sensing capabilities of ISAC, the practical implementation aspect and the resulting overhead reduction remain less explored. Consequently, further investigation is necessary to quantify the actual amount of overhead reduction achievable in practical systems when implementing ISAC techniques within the NR-V2X framework.

To address the aforementioned challenges, we propose innovative frame structures and transmission protocols for 5G NR that enable sensing-assisted beam management in V2I networks, aiming to minimize the overhead caused by pilot and reference signals. The feasibility of the proposed approach is evaluated through link-level simulations, employing real channel conditions \cite{3gpp.38.901} between building groups and vehicles to showcase its enhanced performance. The main contributions of this paper are summarized as follows:\looseness=-1 

\begin{itemize}
    \item \textbf{ISAC-based frame structures for initial access and connected mode.} Building upon the existing frame structures of 5G NR, we investigate the pilot and reference signals, identifying opportunities to eliminate redundant components with the aid of ISAC technologies, and thereby reduce overhead. The proposed ISAC-based frame structures maintain the fundamental functionalities of the system while improving communication performance through optimized resource allocation.\looseness=-1 
    \item \textbf{Kinematic parameters-based fast beam failure detection and recovery.} In order to mitigate latency associated with conventional beam failure detection and recovery mechanisms, we propose an algorithm that detects and recovers from beam failures based on abrupt changes in the kinematic parameters of the vehicle targets. By continuously monitoring beam failure instead of relying solely on periodic reference signals, the proposed algorithm significantly reduces the time required for beam failure detection and recovery, addressing latency concerns.\looseness=-1 
\end{itemize}

The remainder of this paper is organized as follows, Section 2 introduces the system models and algorithms used in sensing-assisted NR-V2I communications, Section 3 describes frame structures comparison between conventional NR and ISAC-based NR in different case studies, Section 4 provides numerical results from link-level simulations in different scenarios, and finally Section 5 concludes the paper.\looseness=-1 

\textbf{Notations:} Without particular specification, matrices,  vectors and scalars are denoted by bold uppercase letters (i.e., $\mathbf{A}$), bold lowercase letters (i.e., $\mathbf{a}$) and normal font (i.e., $N$), respectively. $\Re(\cdot)$ and $\Im(\cdot)$ are respectively used to denote the real and imaginary parts of a complex number. $(\cdot)^*$, $(\cdot)^T$, $(\cdot)^H$ and $(\cdot)^{-1}$ represent the conjugate, transpose, Hermitian and inverse operators, respectively. $\otimes$ and $\odot$ denote the tensor product and the Hadamard product. Moreover, $\mathcal{N}(\mu, \sigma^2)$ and $\mathcal{CN}(\bm{\mu}, \mathbf{R})$ denote the Gaussian distribution with mean $\mu$ and variance $\sigma^2$, and the complex Gaussian distribution with mean $\bm{\mu}$ and covariance matrix $\mathbf{R}$. \looseness=-1 

\section{Sensing-assisted Communications in NR-V2I Networks}
\begin{figure}[htbp]
\centering
\includegraphics[width=\columnwidth]{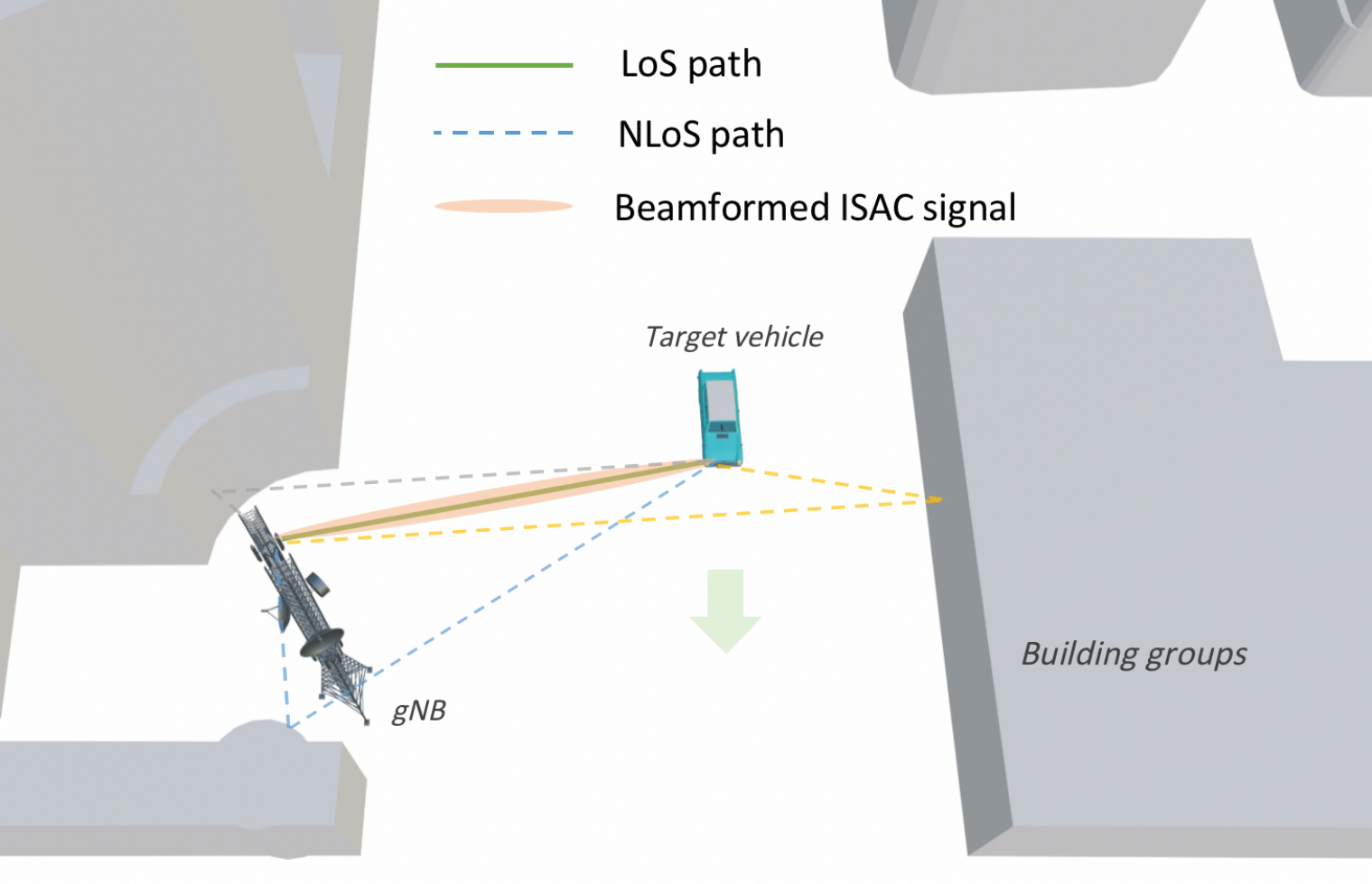}
\caption{Simulation scenario}
\label{scenario}
\end{figure}

In this paper, we investigate the V2I network operating within the framework of the 5G NR protocol, specifically in the mmWave frequency band, which adopts OFDM waveform for data transmission. We assume that the gNB operates in full-duplex mode and is equipped with an mMIMO uniform planar array (UPA) which has $N_t$ transmit antennas and $N_r$ receive antennas. The vehicle in the network is also equipped with a MIMO UPA and is driving down the road while maintaining continuous communication with the gNB. The communication link between the gNB and the vehicle consists of one LoS path and $K-1$ NLoS paths, as shown in Fig. \ref{scenario}.

The maximum time duration of interest is denoted as $T_{\text{max}}$ and can be discretized into multiple small time slots with duration of $\Delta T$, each of which is shorter than the coherence time of the communication channel. Then, it is reasonable to assume that all relevant parameters remain constant within each time slot. All the parameters are defined in the time window of $t \in[0, T_{\text{max}}]$. Without loss of generality, the kinematic parameters of the vehicle at the $n$th time slot are denoted as $\bm{\theta}_{n}=[\theta_{n}, \phi_{n}]^T$, $d_n$ and $v_n$, which represent the azimuth and elevation angles, range and speed of the vehicle, respectively. 

\subsection{Radar Signal Model}

Let us denote the transmitted OFDM signal at the $n$th time slot with $M$ subcarriers in frequency domain and $L$ symbols in time domain as 
\begin{equation} \label{1}
\begin{aligned}
s_n(t)=\sum_{l=0}^{L-1}\sum_{m=0}^{M-1} s_{m,l} e^{j 2 \pi m \Delta f t} \operatorname{rect}(\frac{t-lT_{\text{s}}}{T_{\text{s}}})
\end{aligned}
\end{equation}
where $\Delta f$ denotes the subcarrier spacing in the OFDM signal and $s_{m,l}$ is the transmitted data carried by the $m$th subcarrier at the $l$th symbol. $T_{\text{s}}=T_{\text{cp}}+T$ denotes the time duration of an OFDM symbol, which is the sum of the time duration of cyclic prefix $T_{\text{cp}}$ and the elementary symbol duration $T$.

The gNB received the reflected echoes from the vehicle and $K-1$ scatterers at the $n$th time slot can be formulated as 
\begin{equation} \label{2}
\begin{aligned}
\mathbf{r}_n(t) = \zeta \sqrt{p}\sum_{k=1}^K  \beta_{k, n}  e^{j 2 \pi \mu_{k, n} t} \mathbf{b}(\bm{\theta}_{k, n}) \mathbf{a}^T(\bm{\theta}_{k, n}) \\
\cdot \mathbf{f}_{n} {s}_n(t-\tau_{k,n}) +\mathbf{r}_{\text{self}}(t) +\mathbf{z}_r(t)
\end{aligned}
\end{equation}
where $\zeta = \sqrt{N_tN_r}$ denotes the array gain factor, with $N_t$ and $N_r$ being the number of transmit and receive antennas respectively, $p$ denotes the transmitted signal power, and $\beta_{k, n}$ and $\mu_{k, n}$ denote the reflection coefficient and the Doppler frequency of the $k$th scatterer, respectively. The reflection coefficient can be expressed as $\beta_{k, n}=\epsilon_{k, n}(2d_{k, n})^{-2}$ if the complex radar cross-section (RCS) $\epsilon_{k, n}$ and the relative distance $d_{k, n}$ are given. The Doppler frequency $\mu_{k, n}=2v_{k,n}f_{c}c^{-1}$ and the time delay $\tau_{k,n}=2d_{k, n}c^{-1}$ are determined by the radial velocity $v_{k,n}$ and the distance $d_{k,n}$ of the $k$th scatterer respectively. $\mathbf{r}_{\text{self}}(t)$ denotes the self-interference incurred by the FD mode. Finally, $\mathbf{z}_r(t)$ denotes the complex additive white Gaussian noise with zero mean.

We assume that gNB's UPA has the inter-element spacing of half-wavelength. Moreover, $\mathbf{a}\left(\bm{\theta}_{k, n}\right)$ and $\mathbf{b}\left(\bm{\theta}_{k, n}\right)$ in (\ref{2}) are the transmit and receive steering vectors of the gNB's UPA respectively, which are in the forms of
\begin{equation}\label{3}
\mathbf{a}(\bm{\theta}_{k, n})=\mathbf{a}(\theta, \phi)=\mathbf{v}_{az}(\theta, \phi) \otimes \mathbf{v}_{el}(\phi)
\end{equation}
\begin{equation}\label{4}
\mathbf{b}(\bm{\theta}_{k, n})=\mathbf{b}(\theta, \phi)=\mathbf{u}_{az}(\theta, \phi) \otimes \mathbf{u}_{el}(\phi)
\end{equation}
where $\theta$ and $\phi$ are the azimuth and elevation angles, and $\mathbf{v}_{az}(\theta, \phi)$, $\mathbf{v}_{el}(\phi)$ and $\mathbf{u}_{az}(\theta, \phi)$, $\mathbf{u}_{el}(\phi)$ are the transmit and receive steering vectors in the horizontal and vertical directions, respectively.
\begin{equation}
\mathbf{v}_{az}(\theta, \phi)=\sqrt{\frac{1}{N_{t,x}}}\left[1, e^{j \pi \sin \theta \cos \phi},\cdots,e^{j \pi\left(N_{t,x}-1\right) \sin \theta \cos \phi}\right]^T
\end{equation}
\begin{equation}
\mathbf{v}_{el}(\phi)=\sqrt{\frac{1}{N_{t,y}}}\left[1, e^{j \pi \sin \phi}, \cdots, e^{j \pi\left(N_{t,y}-1\right) \sin \phi}\right]^T
\end{equation}
\begin{equation}
\mathbf{u}_{az}(\theta, \phi)=\sqrt{\frac{1}{N_{r,x}}}\left[1, e^{j \pi \sin \theta \cos \phi}, \cdots, e^{j \pi\left(N_{r,x}-1\right) \sin \theta \cos \phi}\right]^T
\end{equation}
\begin{equation}
\mathbf{u}_{el}(\phi)=\sqrt{\frac{1}{N_{r,y}}}\left[1, e^{j \pi \sin \phi}, \cdots, e^{j \pi\left(N_{r,y}-1\right) \sin \phi}\right]^T
\end{equation}
where $N_{t,x}$, $N_{t,y}$ and $N_{r,x}$, $N_{r,y}$ denote the number of transmit and receive antennas in each row and column of the UPA respectively. The beamforming vector at the $n$th slot $\mathbf{f}_{n}$ is designed based on the predicted angle $\hat{\bm{\theta}}_{n | n-1}$ from the $\left(n-1\right)$th slot as
\begin{equation}
\mathbf{f}_{n}=\mathbf{a}(\hat{\bm{\theta}}_{n | n-1})
\end{equation}

\subsection{Radar Measurement Model}
By sampling at each OFDM symbol and performing block-wise Fourier Transform, the signal can be discretized and presented in the frequency domain. Then, at the $n$th time slot, the received discrete signal at the $i$th antenna and the $l$th symbol can be expressed as :
\begin{equation}
\begin{aligned}
\mathbf{r}_{i,l} = \zeta \sqrt{p}\sum_{k=1}^K \beta_{k} [\mathbf{b}(\bm{\theta}_{k,n})]_i \mathbf{a}^T(\bm{\theta}_{k,n}) \mathbf{f}_n \mathbf{D}(\mu_{k,n}) \\
\cdot \left( \mathbf{s}_{l} \odot \bm{\eta}(\tau_{k,n})[\bm{\omega}^*(\mu_{k,n})]_{l} \right) + \mathbf{z}_{i,l}
\end{aligned}
\end{equation}
where $i=0,...,N_r-1$, $\mathbf{r}_{i,l}=[r_{i,l}[0],.,r_{i,l}[M-1]]^T$, $\mathbf{s}_{l}=[s_{0,l},.,s_{M-1,l}]^T$, and
\begin{equation}
\begin{aligned}
\bm{\eta}(\tau_{k,n}) = \left[1, e^{-j 2 \pi \Delta f \tau_{k,n}}, \cdot, e^{-j 2 \pi \Delta f \left(M-1\right) \tau_{k,n}}\right]^T
\end{aligned}
\end{equation}
\begin{equation}
\begin{aligned}
\bm{\omega}(\mu_{k,n}) = \left[1, e^{-j 2 \pi \mu_{k,n} T_{s}}, \cdot, e^{-j 2 \pi \mu_{k,n} \left(L-1\right) T_{s}}\right]^T
\end{aligned}
\end{equation}
\begin{equation}
\begin{aligned}
\mathbf{D}(\mu_{k,n}) = \operatorname{diag}\left(1, e^{j 2 \pi \mu_{k,n} \frac{T_{s}}{M}}, \cdot, e^{j 2 \pi \mu_{k,n} \frac{(M-1)T_{s}}{M}}\right)
\end{aligned}
\end{equation}
$\mathbf{z}_{i,l}$ is the additive Gaussian noise sample. The diagonal matrix $\mathbf{D}(\mu_{k,n})$ denotes the inter-carrier interference (ICI) in the fast-time domain. We assume that the time duration of the OFDM symbol $T_{\text{s}}$ is significantly smaller than the coherence time, which is inversely proportional to the Doppler shift, leading to $\mu_{k,n} T_{\text{s}} \ll 1$. The self-interference term is neglected for simplicity, assuming that any observed Doppler shift is solely due to the motion of the target and not influenced by the gNB's own transmitted signal. Thus, the ICI matrix exhibits proximity to the identity matrix. By aggregating $L$ symbols in each time slot, the radar signal received at the $i$th antenna can be written into a compact matrix form:
\begin{equation} 
\begin{aligned}
\mathbf{R}_{i} = \sum_{k=1}^K \alpha_{k} \left( \mathbf{S} \odot \bm{\eta}(\tau_{k,n}) \bm{\omega}^H(\mu_{k,n}) \right) + \mathbf{Z}_{i}
\end{aligned}
\end{equation}
where $\alpha_{k}=\zeta \sqrt{p}\beta_{k} [\mathbf{b}(\bm{\theta}_{k,n})]_i \mathbf{a}^T(\bm{\theta}_{k,n}) \mathbf{f}_n$, $\mathbf{R}_{i}=[\mathbf{r}_{i,0},.,\mathbf{r}_{i,L-1}]\in\mathbb{C}^{M \times L}$ and $\mathbf{S}=[\mathbf{s}_{0},.,\mathbf{s}_{L-1}]\in\mathbb{C}^{M \times L}$. $\mathbf{Z}_{i} \in \mathbb{C}^{M \times L}$ denotes the additive Gaussian noise matrix where $\operatorname{vec}(\mathbf{Z}_{i})\sim \mathcal{CN}(\mathbf{0}_{ML}, \sigma^2\mathbf{I}_{ML})$.
The channel information can be extracted independently from the payload data by doing element-wise division between the received signal and the transmitted signal\cite{sturm2011waveform}. The post-processed signal matrix at the $i$th antenna can be written as:
\begin{equation} \label{15}
\begin{aligned}
\mathbf{\Tilde{R}}_{i} &= \sum_{k=1}^K  \alpha_{k} \bm{\eta}_{\tau_{k,n}} \bm{\omega}_{\mu_{k,n}}^H + \mathbf{\Tilde{Z}}_{i} \\
&= \mathbf{X}_{i}+\mathbf{\Tilde{Z}}_{i}
\end{aligned}
\end{equation}
where $\operatorname{vec}(\Tilde{\mathbf{Z}}_{i})\sim \mathcal{CN}(\mathbf{0}_{ML}, \Tilde{\sigma}^2\mathbf{I}_{ML \times 1})$ with $\Tilde{\sigma}^2 = \frac{\sigma^2}{ML}\operatorname{Tr}\left( \mathbf{\Tilde{S}}^{-1} \mathbf{\Tilde{S}}^{-H} \right)$ and $\mathbf{\Tilde{S}}= \operatorname{diag}\left(\operatorname{vec}(\mathbf{S})\right)$.
The channel information in (\ref{15}) consists of the time delay $\tau_{k,n}$ and Doppler frequency $\mu_{k, n}$, which can be effectively leveraged for the estimation of the key target parameters, including range and velocity. Thus, to improve the precision and reduce the complexity of the estimation, we propose a two-step algorithm to extract the range and velocity of the target. 

First, by implementing 2D-DFT with respect to $\mathbf{\Tilde{R}}_{i}$, we may get a rough estimation of the range and velocity. To be specific, by employing IFFT and FFT on the fast-time domain and slow-time domain respectively, scatterers and the target are detected at each corresponding index $(\hat{m}_k,\hat{l}_k)$. The resulting distance and radial velocity can be given as
\begin{equation}
\begin{aligned}
\hat{d}_{k}=\frac{\hat{m}_k c}{2 M \Delta f}
\end{aligned}
\end{equation}
\begin{equation}
\begin{aligned}
\hat{v}_{k}=\frac{\hat{l}_k c}{2 f_c L T_{\text{s}}}
\end{aligned}
\end{equation}
where $f_c$ and $c$ denote the carrier frequency and the speed of light respectively. The scatterers are assumed to be generated from static targets, e.g., buildings. Consequently, the parameters of the moving vehicles can be extracted from the detected peaks by simply eliminating peaks with zero relative velocities.  

Subsequently, the MUSIC algorithm \cite{braun2011single} is employed to achieve super-resolution estimation of the range and velocity of the target. Specifically, we identify $(\Tilde{m},\Tilde{l})$ as the corresponding peak index of the target through 2D-DFT. To mitigate computational complexity and reduce time cost, a much narrower interval can be applied in the MUSIC algorithm. In contrast to the conventional grid-based techniques commonly employed in MUSIC, we leverage the Golden-section search \cite{Kiefer1953} to expedite convergence and enhance estimation accuracy. The radian frequency of the range and velocity can be defined as $w_{d}=2 \pi \Delta f \tau$ and $w_{v}=2 \pi T_{s} \mu$ respectively\cite{xie2021performance}. Accordingly, the steering vectors of them are expressed as
\begin{equation}
\begin{aligned}
\bm{a}_{d} = \left[1, e^{-j w_{d}}, \cdot, e^{-j \left(M-1\right) w_{d}}\right]^T
\end{aligned}
\end{equation}
\begin{equation}
\begin{aligned}
\bm{a}_{v} = \left[1, e^{j w_{v}}, \cdot, e^{j \left(L-1\right) w_{v}}\right]^T
\end{aligned}
\end{equation}
 Following the steps of Golden-section search, the initial searching points of $\tau$ and $\mu$ would be $[\tau_{a},(1-\chi)(\tau_{b}-\tau_{a}),\chi(\tau_{b}-\tau_{a}),\tau_{b}]$ and $[\mu_{a},(1-\chi)(\mu_{b}-\mu_{a}),\chi(\mu_{b}-\mu_{a}),\mu_{b}]$ respectively, where $\chi=\frac{\sqrt{5}-1}{2}$, $\tau_{a}=\frac{\Tilde{m}-1}{M\Delta f }$, $\tau_{b}=\frac{\Tilde{m}+1}{M\Delta f }$, $\mu_{a}=\frac{\Tilde{l}-1}{LT_{\text{s}} }$, $\mu_{b}=\frac{\Tilde{l}+1}{LT_{\text{s}} }$. To estimate the range, the MUSIC algorithm can be applied to the processed received signal at the $l$th symbol:
\begin{equation}
\begin{aligned}
\mathbf{\Tilde{R}}_{l}=[\mathbf{\Tilde{r}}_1, ...,\mathbf{\Tilde{r}}_{N_r-1}] \in \mathbb{C}^{M \times N_r}
\end{aligned}
\end{equation}
The covariance matrix can be written as $\bm{\Sigma_{\tau}}=\frac{1}{L}\sum_{l=1}^L\mathbf{\Tilde{R}}_{l}\mathbf{\Tilde{R}}_{l}^H$. After eigenvalue decomposition, it can be further expressed as:
\begin{equation}
\begin{aligned}
\bm{\Sigma_{\tau}}=\mathbf{U}_s \bm{\Lambda}_s \mathbf{U}_s^H+\mathbf{U}_n \bm{\Lambda}_n \mathbf{U}_n^H
\end{aligned}
\end{equation}
where the diagonal matrices $\bm{\Lambda}_s$ and $\bm{\Lambda}_n$ each contains $K$ and $M-K$ eigenvalues, and the signal and noise subspace $\mathbf{U}_s$ and $\mathbf{U}_n$ contains $K$ and $N_r-K$ eigenvectors. The MUSIC spectrum can be yielded by 
\begin{equation}
\begin{aligned}
P_{\text{MUSIC}}(\tau)=\frac{1}{\bm{{a}_{d}}^{H}(\tau) \mathbf{U}_n \mathbf{U}_n^H \bm{{a}_{d}}(\tau)} 
\end{aligned}
\end{equation}
By narrowing the interval of the searching in each iteration, the peak can be determined to be the mid point of the interval once the interval is smaller than a certain threshold. The estimation of velocity follows the same procedure whereas it has different covariance matrix $\bm{\Sigma_{\mu}}=\frac{1}{M}\sum_{m=1}^M\mathbf{\Tilde{R}}_{m}\mathbf{\Tilde{R}}_{m}^H$ and MUSIC spectrum $P_{\text{MUSIC}}(\mu)$, where $\mathbf{\Tilde{R}}_{m}$ denotes the processed received signal at the $m$th subcarrier and can be expressed as:
\begin{equation}
\begin{aligned}
\mathbf{\Tilde{R}}_{m}=[\mathbf{\Tilde{r}}_1, ...,\mathbf{\Tilde{r}}_{N_r-1}] \in \mathbb{C}^{L \times N_r}
\end{aligned}
\end{equation}
The detailed steps of the range and velocity estimation can be summarized in Algorithm \ref{alg1}.

\begin{algorithm}
\caption{Range and Velocity estimation}\label{alg1}
\begin{algorithmic}
\Require $\tau_{a},\tau_{b},\mu_{a},\mu_{b}$
\Ensure $\tau, \mu$
\State Threshold of the interval of $\tau$: $\tau_{\text{thre}}$
\State Threshold of the interval of $\mu$: $\mu_{\text{thre}}$
\While{$\tau_{a}-\tau_{b}\geq\tau_{\text{thre}}$}
  \If{$P_{\text{MUSIC}}((1-\chi)(\tau_{b}-\tau_{a})) \geq P_{\text{MUSIC}}(\chi(\tau_{b}-\tau_{a}))$}
      \State $\tau_{b} \gets \chi(\tau_{b}-\tau_{a})$
  \Else
      \State $\tau_{a} \gets (1-\chi)(\tau_{b}-\tau_{a})$
  \EndIf
\EndWhile
\State $\tau=\frac{\tau_{a}+\tau_{b}}{2}$
\State By doing similar procedures, $\mu=\frac{\mu_{a}+\mu_{b}}{2}$
\end{algorithmic}
\end{algorithm}
The resulting range and radial velocity of the target can be expressed as
\begin{equation} \label{24}
\begin{aligned}
d=\frac{\tau c}{2}, v=\frac{\mu c}{2 f_c}
\end{aligned}
\end{equation}

Furthermore, 2D-MUSIC can be employed to estimate the DOA of the target with the covariance matrix to be $\bm{\Sigma_{\bm{\theta}}}=\frac{1}{L}\sum_{l=1}^L\mathbf{\Tilde{R}}_{l}^H\mathbf{\Tilde{R}}_{l}$.
By traversing all the possible directions of the receiving steering vector $\mathbf{b}(\bm{\theta})$ in (\ref{4}), the peak of the MUSIC spectrum $P_{\text{MUSIC}}(\bm{\theta})$ can be detected, thus the angle of the target can be estimated. Although the reflection coefficient is not measured in a direct manner, it can be calculated based on the measurement of $d_{k,n}$.

\subsection{Communication Signal Model}
At the $n$th time slot, the signal received by the vehicle side from the gNB can be formulated as
\begin{equation}
\begin{aligned}
c_n(t)=  \Tilde{\zeta} \sqrt{p}\sum_{k=1}^K  \Tilde{\alpha}_{k, n} \mathbf{v}_n^T \mathbf{u}(\bm{\theta}_{k, n}) \mathbf{a}^T(\bm{\theta}_{k, n}) \mathbf{f}_{n} {s}_{n}(t) +z_c(t)
\end{aligned}
\end{equation}
where $\Tilde{\zeta} = \sqrt{N_tM_r}$ denotes the array gain factor, with $M_r$ being the number of receive antennas at the vehicle side and $\Tilde{\alpha}_{k, n}$ are the path-loss coefficients of different paths.  The receive steering vector $\mathbf{u}\left(\bm{\theta}_{k, n}\right)$ has the similar expression as (\ref{3}). The vehicle side's receive beamforming vector is derived based on the two-step prediction in \cite{liu2020radar}, given as 
\begin{equation}
\mathbf{v}_{n}=\mathbf{u}(\hat{\bm{\theta}}_{n | n-2})
\end{equation}
Here, we define the receive Signal-to-Noise Ratio (SNR) as
\begin{equation}
\begin{aligned}
\operatorname{SNR}_r= \frac{p\left|\Tilde{\zeta}\sum_{k=1}^K  \Tilde{\alpha}_{k, n} \mathbf{v}_n^T \mathbf{u}(\bm{\theta}_{k, n}) \mathbf{a}^T(\bm{\theta}_{k, n}) \mathbf{f}_{n}\right|^2}{{\sigma_c}^2} 
\end{aligned}
\end{equation}
where ${\sigma_c}^2$ denotes the variances of the white Gaussian noise.

\section{Frame Structures and Case studies at mmWave Frequency Band in V2I Networks}
In this section, we introduce the conventional communication-only and the proposed ISAC-based frame structures of NR at mmWave frequencies, and provide case studies of three essential stages in NR beam management to analyze the superiority of sensing-assisted communications in V2I network.

\subsection{General Frame Structure and Reference Signals in 5G NR}
Similar to LTE, OFDM waveform is still adopted in NR. But unlike LTE with only one type of numerology, up to 7 frame structures with different numerologies $\mu$ are supported in NR, which is defined by 3GPP specification \cite{3gpp.38.211}. The relationship between subcarrier spacing and numerology can be expressed as $\Delta f = 2^{\mu} \cdot 15$, $\mu\in\mathbb{N}$, $\mu\leq6$, where $\mu\in[2,6]$ are used in the mmWave frequency band. The time duration of one radio frame and one subframe are 10\text{ms} and 1\text{ms} regardless of the numerology, with each subframe further divided into $2^{\mu}$ slots. Each slot is comprised of either 14 symbols or 12 symbols in case of normal CP or extended CP. The smallest physical resource in NR is named as resource element (RE), which occupies one subcarrier in frequency domain and one symbol in time domain. Each resource block (RB) contains 12 subcarriers in frequency domain and the whole frequency band is made up by multiple RBs. Moreover, up to 61 slot formats for normal CP are supported in Time Division Duplex (TDD) to assign the function at symbol level (i.e., Downlink, Uplink or Flexible) in each time slot. 

\begin{figure}[htbp]
\centering
\includegraphics[width=\columnwidth]{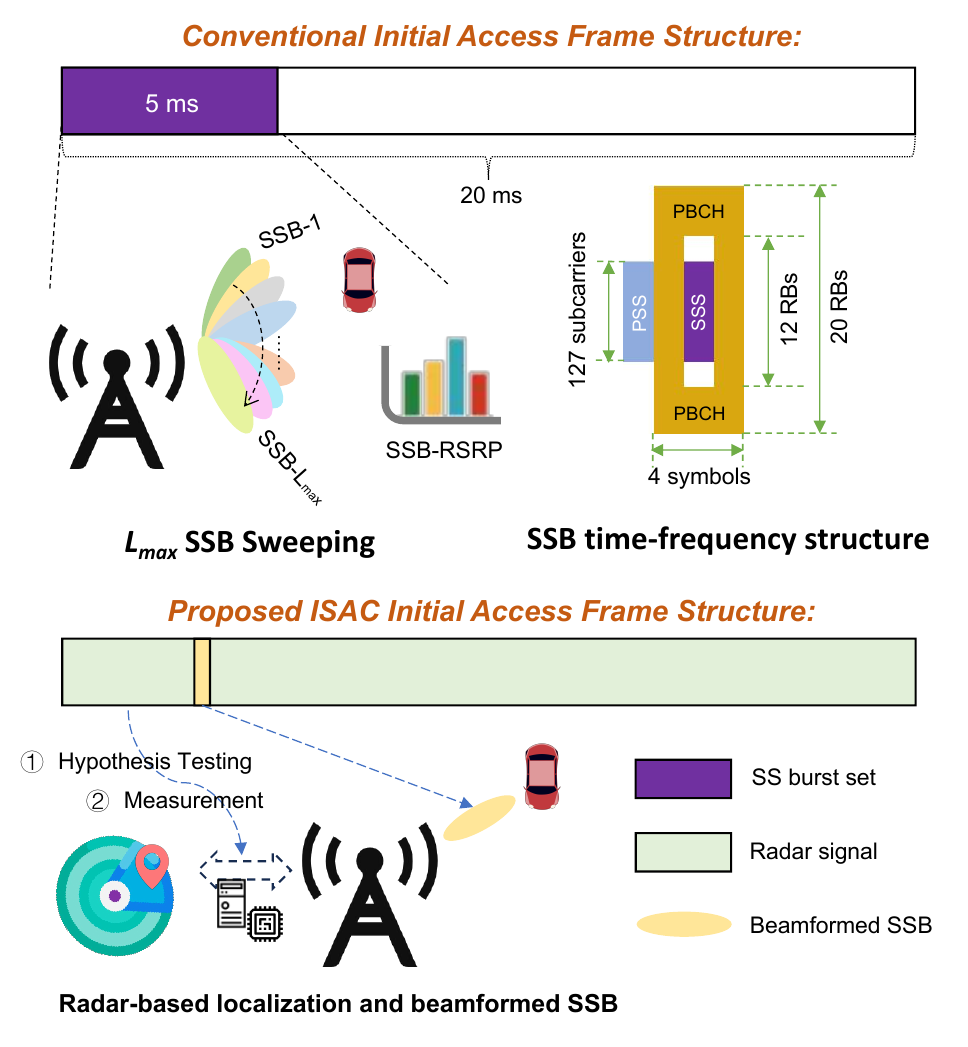}
\caption{Comparison between conventional communication-only frame structure and proposed ISAC-based frame structure in initial access.}
\label{Initial access}
\end{figure}

In 5G NR, reference signals are essential in facilitating various aspects of wireless communications, including channel estimation, beamforming and synchronization. Some key types of downlink reference signals in 5G NR are:

\begin{itemize}
    \item {\it Synchronization Signal Block (SSB):} An SSB consists both synchronization signals and physical broadcast channel (PBCH). Synchronization signals are composed of primary synchronization signal (PSS) and secondary synchronization signal (SSS), which are both aimed to assist initial cell identifications. PBCH demodulation reference signals (DMRS) and PBCH data constitute the PBCH that carries the system information and user data. An SSB occupies 4 symbols and 240 subcarriers in time domain and frequency domain respectively, which is presented in Fig. \ref{Initial access}. 
    \item {\it Demodulation Reference Signal (DMRS):} DMRS plays a key function in the coherent demodulation of PDSCH. It provides the necessary reference for accurate demodulation of the received data symbols. Various mapping types, density options and additional DMRS configurations are supported to cater to different system requirements. 
    \item {\it Channel State Information Reference Signal (CSI-RS):} CSI-RS is utilized for acquiring downlink channel state information and beam refinement. Up to 32 ports are supported in NR, providing options for multiple antenna configurations. The choice of codebook depends on factors such as the number of ports, panel type and the number of users in MIMO systems, enabling adaptive transmission strategies in diverse network scenarios. Upon receiving the CSI-RS, UE processes the signal and extracts crucial parameters which form the basis for reporting back to the gNB. Through CSI-RS feedback, the report that sends back to gNB from UE contains parameters like rank indicator (RI), precoding matrix indicator (PMI), and channel quality information (CQI). The PMI allows the UE to give a recommendation of the preferred precoding matrix for the downlink transmission, which helps in beam refinement and beam switching. The configuration of the CSI-RS in time domain can be periodic, aperiodic or semi-persistent, which depends on its usage and intended purpose. For channel estimation and channel quality monitoring, the period of CSI-RS transmission and its corresponding feedback are determined from a set of discrete values, denoted as $T_{\text{CSI-RS}}\in \{4, 5, 8, 10, 16, 20, 32, 40, 64, 80, 160, 320, 640\}$ slots. 
    \item {\it Phase Tracking Reference Signal (PTRS):} To compensate for the common phase error caused by phase noise generated in local oscillators, PTRS is introduced. It assists in mitigating the adverse effects of phase noise on the received signal and improves the accuracy of phase tracking.
\end{itemize}

\subsection{Initial Access}
Initial access (IA) plays a critical role in the NR protocol where the idle users initialize connections with the network through establishing a reliable communication link with the gNB\cite{3gpp.38.300}. This process aims for realizing the downlink and uplink synchronization and assigning users a specified ID for the upcoming communication. After that, a beam refinement process is required between the gNB and the user, which leads to better signaling quality via optimized beamforming performance.

\subsubsection{IA in conventional NR}
IA in 5G NR can be summarized into three stages, beam sweeping, beam measurement and determination, beam reporting.

\begin{itemize}[leftmargin=*]
    \item {\it Beam sweeping:} Beam sweeping is achieved by transmitting multiple SSBs in downlink direction. An SS burst set comprises $L_{\max}$ SSBs, where $L_{\max}\in \{4,8,64\}$ and varies under different numerologies and frequency bands \cite{3gpp.38.213}. Each SSB in the SS burst set is beamformed towards a certain angle so that the whole set could sweep the coverage area in both azimuth and elevation directions. The period of 20\text{ms} can be assumed by the user for initial cell search. Although up to 64 SSBs are supported in an SS burst set, all of them are transmitted in the first 5\text{ms} of the 20\text{ms} period.
    
    \item {\it Beam measurement and determination:} The user determines the best SSB beam by measuring the SS reference signal received power (SS-RSRP), which is defined as the linear average over the power contributions of the resource elements that carry SSS\cite{3gpp.38.215}. 
\begin{equation}
\begin{aligned}
\text{SS-RSRP (in dBm)} = 10\log_{10}\left(\frac{1}{N} \sum_{n=1}^N \left| \mathbf{X}[n] \right|^2\right)+30
\end{aligned}
\end{equation}
where $\mathbf{X}[n]$ and $N$ denote the SSS resource element and its number, respectively. Meanwhile, downlink synchronization is realized to estimate the time and frequency offset by performing matched filtering. PBCH decoding can be done after extracting the useful system information.

   \item {\it Beam reporting:} The feedback of the best SSB beam is transmitted uplink in the random access channel (RACH). After receiving the RACH preamble, random access response (RAR) is beamformed and transmitted downlink in the direction of the best SSB beam. As the user receives the RAR, beam establishment and synchronization in IA are assumed to be completed.
\end{itemize}

\subsubsection{IA in ISAC NR}
The beam sweeping procedure in conventional communication-only NR IA switches beams to cover each subsection in the coverage area, leading to substantial time delay of connection. In contrast, the sensing ability in ISAC signal could provide real-time monitoring on the movement of the vehicles and initialize synchronization and tracking as soon as the target enters the coverage area of the gNB.

To be more specific, the gNB leverages the omnidirectional radar signal to detect if there exists new targets entering the coverage area when the communication is in idle mode. The implementation of 2D-DFT to the channel transfer information $\mathbf{\Tilde{R}}_{i}$ in (\ref{15}) displays the received signal in delay-Doppler domain as:
\begin{equation}
\begin{aligned}
\mathbf{Y}_{i} &= \mathbf{F}^H_{M}\mathbf{\Tilde{R}}_{i}\mathbf{F}_{L} \\
 &= \mathbf{F}^H_{M}\mathbf{X}_{i}\mathbf{F}_{L} + \mathbf{F}^H_{M}\mathbf{\Tilde{Z}}_{i}\mathbf{F}_{L}
\end{aligned}
\end{equation}
where $\mathbf{F}_{L} \in \mathbb{C}^{L \times L}$ and $\mathbf{F}^H_{M} \in \mathbb{C}^{M \times M}$ denote the $L$-point DFT matrix and $M$-point IDFT matrix, respectively. Following the derivation in \cite{ehsanfarhypothesis}, the target presence detection problem at the $i$th antenna, the $m$th subcarrier and the $l$th symbol can be formulated into a binary hypothesis testing as:
\begin{equation}
\begin{aligned}
\left\{\begin{array}{l}
\mathcal{H}_0: \Re \left(\mathbf{Y}_{i,m,l}\right)=\Re\left(\mathbf{F}^H_{M}\mathbf{\Tilde{Z}}_{i}\mathbf{F}_{L}\right)_{m, l} \\
\mathcal{H}_1: \Re\left(\mathbf{Y}_{i,m,l}\right)=\Re\left(\mathbf{F}^H_{M}\mathbf{X}_{i}\mathbf{F}_{L} + \mathbf{F}^H_{M}\mathbf{\Tilde{Z}}_{i}\mathbf{F}_{L}\right)_{m, l}
\end{array}\right.
\end{aligned}
\end{equation}
which is further simplified as:
\begin{equation}
\begin{aligned}
\left\{\begin{array}{l}
\mathcal{H}_0: \Re\left(\mathbf{Y}_{i,m,l}\right) \sim \mathcal{N}\left(0, \Tilde{\sigma}^2 / 2\right)   \\
\mathcal{H}_1: \Re\left(\mathbf{Y}_{i,m,l}\right) \sim \mathcal{N}\left(\mu, \Tilde{\sigma}^2 / 2\right)   
\end{array}\right.
\end{aligned}
\end{equation}
where $\mu > 0$, $\mathcal{H}_0$ and $\mathcal{H}_1$ denote the hypothesis of the vehicle being absent and present, respectively. The threshold determined by the false alarm rate $P_{\mathrm{FA}}$ is:
\begin{equation}
\begin{aligned}
\left|\Re\left(\mathbf{Y}_{i,m,l}\right)\right| \stackrel{\mathcal{H}_1}{\underset{\mathcal{H}_0}{\gtrless}} \sqrt{\frac{\Tilde{\sigma}^2}{2}} Q^{-1}\left(P_{\mathrm{FA}}\right)
\end{aligned}
\end{equation}

Once the angle and the location of the target are being detected, the gNB sends the beamformed SSB signal in the specific direction and waits for the response to complete the IA procedure. Therefore, instead of transmitting 64 beams in mmWave frequency band every 20ms, only one beamformed synchronization signal will be sent to minimize the time consumed by IA procedure. In the event of a potential failure by the gNB to detect the presence of the target, we consider a scenario where, if a vehicle remains unconnected within the gNB coverage area for a specified duration, such as 20\text{ms}, without receiving the SSB signal, the vehicle initiates an uplink signal transmission to the gNB. Following this, the gNB proceeds to execute the conventional communication-only IA procedure.

\begin{figure*}[htbp]
\centering
\includegraphics[width=18cm]{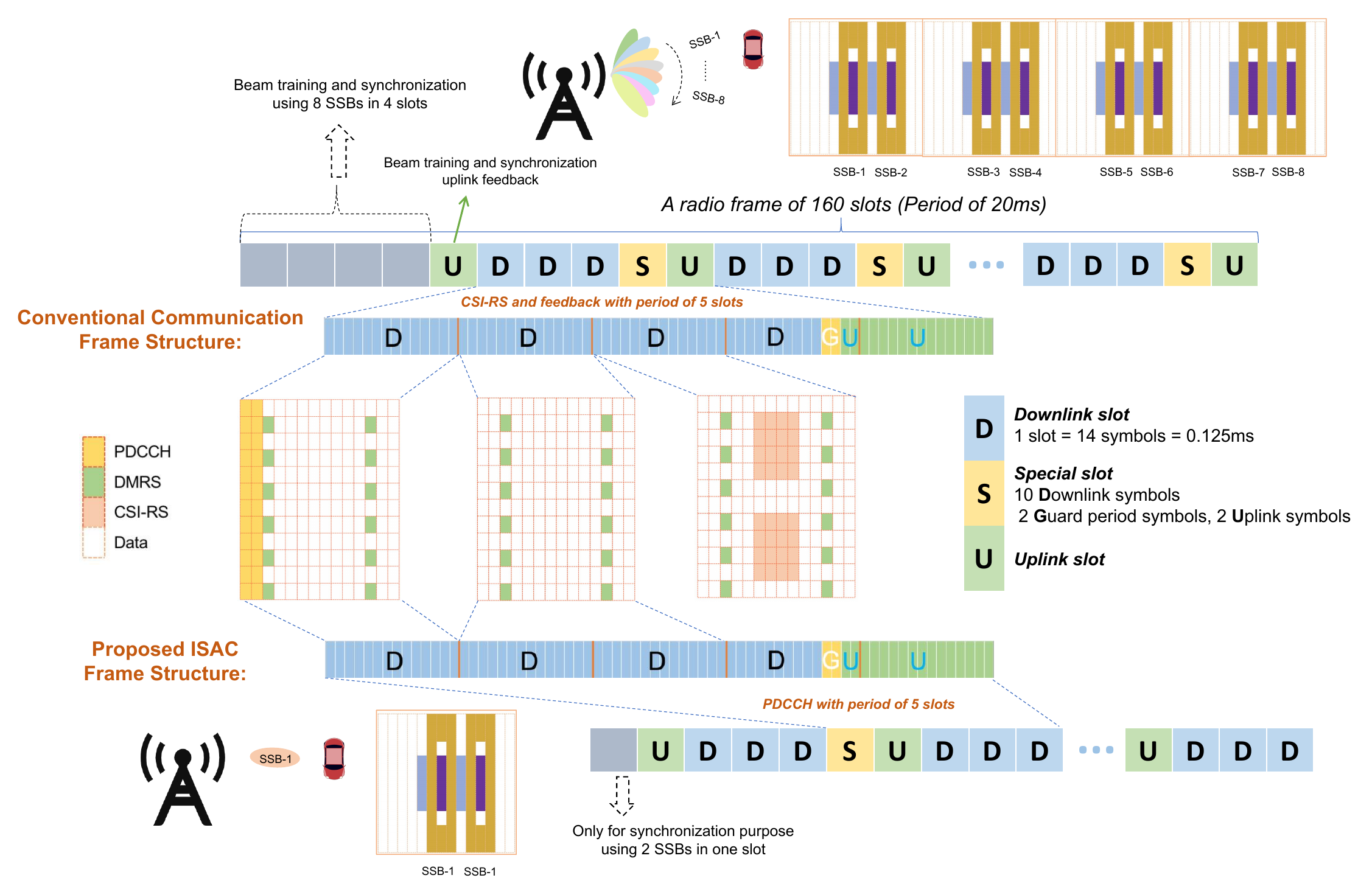}
\caption{Comparison between conventional communication frame structure and proposed ISAC frame structure in connected mode}
\label{connected}
\end{figure*}

\subsection{Connected Mode}

After undergoing the IA procedure, the UE establishes and maintains a connection with the gNB, a state commonly referred to as the connected mode\cite{3gpp.38.331}\cite{3gpp.38.804}. Connected mode enables the UE to access various services provided by the V2I network, such as real-time data transmission of surrounding traffic and environment. In the connected mode, the UE and the gNB in the V2I network engage in ongoing interactions, including exchange of control information, synchronization, handover procedures and quality-of-service (QoS) management. To ensure the reliable and efficient communication between the UE and the gNB, the minimization of pilot signals and reference signals is vital in connected mode to maintain high-speed data transmission. However, it is important to acknowledge that reducing the usage of pilot and reference signals can potentially lead to an increased likelihood of errors in demodulation or beam misalignment. Fortunately, the sensing ability in ISAC V2I systems offers novel opportunities to enhance the efficiency of beam tracking procedures while minimizing the need for excessive pilot and reference signals.

\subsubsection{Connected Mode in conventional NR}
In the connected mode of NR, maintaining precise match filtering and coherent demodulation by the UE necessitates synchronization with the gNB. This synchronization is achieved through periodic transmission of synchronization reference signals, specifically SSBs. Furthermore, in the connected mode, the periodic transmission of SSBs may also be considered as a form of beam training. This process enables accurate beam alignment between the UE and the gNB, which is vital for enhancing overall communication quality. To reduce the overhead of beam training, only a subset of the SSBs is transmitted periodically in the connected mode. Moreover, the period of the SSBs transmission $T_{\text{SS}}$ has more options instead of the fixed period of 20\text{ms} in initial access, where $T_{\text{SS}}=2^{k}\cdot5$\text{ms}, $k\in\mathbb{N}$, $k\leq5$. 

Subsequently, in preparation for data transmission, Downlink Control Information (DCI) which provides essential scheduling information regarding the allocation of transmission resources to the UE, is conveyed through the Physical Downlink Control Channel (PDCCH). During the actual data transmission, the primary carrier of the transmitted data is the Physical Downlink Shared Channel (PDSCH). Within the PDSCH, multiple types of reference signals are mapped to aid in various aspects of reception\cite{3gpp.38.211}, such as DMRS, CSI-RS and PTRS.

In this study, we consider a specific single layer SU-MIMO V2I network under the 5G NR framework with the chosen numerology denoted as $\mu=3$, signifying the system is operating under mmWave frequency band with the subcarrier spacing of 120kHz. The period of SSBs transmission is still considered to be 20ms and only a subset of 8 SSB beams are transmitted instead of the whole 64 SSB beams. We adopt a widely used 5G NR frame structure, denoted as ``$DDDSU$'' in V2I network, where ``$D$'', ``$S$'' and ``$U$'' represent the downlink slot, special slot and uplink slot respectively\cite{ATIS.3GPP.37.910.V1610}.  Given the high mobility of the vehicles in the V2I network, we assume that the DMRS in the PDSCH adopts the mapping type of ``$A$'' and an additional DMRS is added. To facilitate CSI estimation, CSI-RS is configured with a periodicity of 5 slots, employing the maximum available 32 antenna ports. The CSI-RS feedback report in the uplink slot contains parameters like CQI and PMI for channel estimation, beam refinement and beam switching, which has the same period as the CSI-RS transmission. For simplicity, PTRS is not considered in this study. The conventional communication-only NR frame structure of the proposed scenario is presented in Fig. \ref{connected}.

\subsubsection{Connected Mode in ISAC NR}

\begin{figure}[htbp]
\centering
\includegraphics[width=8cm]{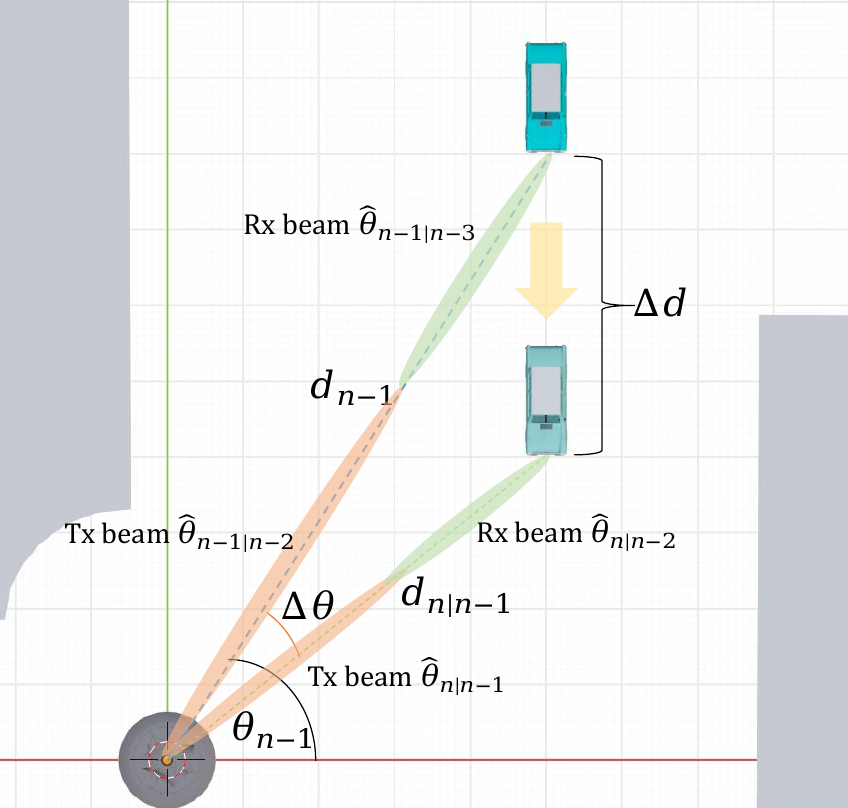}
\caption{State Evolution Model}
\label{state_evolution_model}
\end{figure}  

Accurate tracking of the vehicle is the crucial element to ensure the high-quality communication in V2I network. To minimize the requirement for excessive pilot signals, it's beneficial to employ an advanced prediction method, such as EKF, for the precise estimation and tracking of the kinematic parameters associated with the vehicle. Following the derivation in \cite{liu2020radar} and based on the geometric relationships in Fig. \ref{state_evolution_model}, the state evolution model can be summarized as
\begin{equation}\label{16}
\left\{\begin{array}{l}
\theta_n=\theta_{n-1}-d_{n-1}^{-1} v_{n-1} \Delta T \cos \theta_{n-1}+\omega_\theta \\
d_n=d_{n-1}-v_{n-1} \Delta T \sin \theta_{n-1}+\omega_d \\
v_n=v_{n-1}+\omega_v \\
\beta_n=\beta_{n-1}\left(1-d_{n-1}^{-1} v_{n-1} \Delta T \sin \theta_{n-1}\right)^2+\omega_\beta
\end{array}\right.
\end{equation}
where $\theta_n$, $d_n$, $v_n$ and $\beta_n$ denote the azimuth angle, distance, velocity and reflection coefficient at the $n$th slot, respectively. The state evolution model and the measurement model derived in (\ref{16}) and (\ref{24}) can be formulated in compact forms as
\begin{equation}
\left\{\begin{array}{l}
\text {State Evolution Model: } \mathbf{x}_n=\mathbf{g}\left(\mathbf{x}_{n-1}\right)+\boldsymbol{\omega}_n \\
\text {Measurement Model: } \mathbf{y}_n=\mathbf{x}_n+\mathbf{z}_n
\end{array}\right.
\end{equation}
where $\mathbf{x}=[\theta, d, v, \beta]^T$ and $\mathbf{y}=[\hat{\theta}, \hat{d}, \hat{v}, \hat{\beta}]^T$ denote the state variable and the measurement variable respectively, $\mathbf{g}$ is defined in (\ref{16}), $\boldsymbol{\omega}=[\omega_\theta, \omega_d, \omega_v, \omega_\beta]^T$ and $\mathbf{z}=[z_\theta, z_d, z_v, z_\beta]^T$ are the zero-mean Gaussian noises caused by approximation and measurement respectively, whose covariance matrices can be expressed as 
\begin{equation}
\mathbf{Q}_s=\operatorname{diag}\left(\sigma_\theta^2, \sigma_d^2, \sigma_v^2, \sigma_\beta^2\right)
\end{equation}
\begin{equation}
\mathbf{Q}_m=\operatorname{diag}\left(\Tilde{\sigma}_\theta^2, \Tilde{\sigma}_d^2, \Tilde{\sigma}_v^2, \Tilde{\sigma}_\beta^2\right)
\end{equation}

The variances of the measurement noises are directly proportional to CRBs from \cite{liu2022survey}. Before performing EKF, the state evolution model $\mathbf{g}$ needs to be linearized by deriving its Jacobian matrix, which can be given as:
\begin{equation}
\begin{aligned}
\frac{\partial \mathbf{g}}{\partial \mathbf{x}}={\left[\begin{array}{cccc}
1+\frac{v \Delta T \sin \theta}{d} & \frac{v \Delta T \cos \theta}{d^2} & -\frac{\Delta T \cos \theta}{d} & 0 \vspace{0.8ex}\\
-v \Delta T \cos \theta & 1 & -\Delta T \sin \theta & 0 \vspace{0.8ex}\\
0 & 0 & 1 & 0 \vspace{0.8ex}\\
-\frac{2 \beta v \Delta T \cos \theta}{d}\iota & \frac{2 \beta v \Delta T \sin \theta}{d^2}\iota & -\frac{2 \beta \Delta T \sin \theta}{d}\iota & \iota^2
\end{array}\right]}
\end{aligned}
\end{equation}
where $\iota=\left(1-\frac{v \Delta T \sin \theta}{d}\right)$. Finally, follow the standard steps of the procedure in \cite{Kay1993}, the prediction and estimation of EKF can be summarized as follows:

{\it {1) State Prediction:}} 
\begin{equation}
\hat{\mathbf{x}}_{n \mid n-1}=\mathbf{g}\left(\hat{\mathbf{x}}_{n-1}\right), \hat{\mathbf{x}}_{n+1 \mid n-1}=\mathbf{g}\left(\hat{\mathbf{x}}_{n \mid n-1}\right).
\end{equation}

{\it {2) Linearization:}} 
\begin{equation}
\mathbf{G}_{n-1}=\left.\frac{\partial \mathbf{g}}{\partial \mathbf{x}}\right|_{\mathbf{x}=\hat{\mathbf{x}}_{n-1}}, \mathbf{H}_n=\mathbf{I}_4
\end{equation}

{\it {3) MSE Matrix Prediction:}} 
\begin{equation}
\mathbf{M}_{n \mid n-1}=\mathbf{G}_{n-1} \mathbf{M}_{n-1} \mathbf{G}_{n-1}^H+\mathbf{Q}_s
\end{equation}

{\it {4) Kalman Gain Calculation:}} 
\begin{equation}
\mathbf{K}_n=\mathbf{M}_{n \mid n-1} \mathbf{H}_n^H\left(\mathbf{Q}_m+\mathbf{H}_n \mathbf{M}_{n \mid n-1} \mathbf{H}_n^H\right)^{-1}
\end{equation}

{\it {5) State Tracking:}} 
\begin{equation}
\hat{\mathbf{x}}_n=\hat{\mathbf{x}}_{n \mid n-1}+\mathbf{K}_n\left(\mathbf{y}_n-\hat{\mathbf{x}}_{n \mid n-1}\right))
\end{equation}

{\it {6) MSE Matrix Update:}} 
\begin{equation}
\mathbf{M}_n=\left(\mathbf{I}-\mathbf{K}_n \mathbf{H}_n\right) \mathbf{M}_{n \mid n-1}
\end{equation}
where $\mathbf{I}_4$ denotes the identity matrix of size four. 

Frequent beam training and reference signals transmission are considered to be overhead in the communication system, due to the fact that the frequency-time resources could be assigned to transmit useful data. This may limit the effectiveness of the communication performance, especially for high-mobility V2I links. Fortunately, with the utilization of ISAC in the NR V2I network, some of the reference signals can be reduced, thus improve the overall throughput, as detailed below. 

The CSI-RS mapped in PDSCH in connected mode is mainly used for channel estimation. To be specific, UE obtains the information of channel based on the received CSI-RS and sends feedback to gNB which contains the parameters that UE prefers, such as PMI and RI. The downlink CSI-RS and the uplink feedback are useful in designing the transmission scheme for the next period but they also cause considerable overheads which will reduce the throughput and data rate. We highlight that, this dilemma can be tackled by applying the ISAC signaling approach in the NR network, which provides the CSI information to the gNB based on the echo signals reflected by the vehicles. The accurate prediction and estimation using EKF provides precise tracking of the target. Therefore, instead of implementing beam training and transmitting reference signals periodically to find best beam pairs and obtain the CSI, we propose to reduce the number of SSBs in transmission and abolish the transmission of CSI-RS. Now that the SSBs only have the function to achieve synchronization, only one SSB is beamformed to the predicted user direction. A repeated one can be transmitted in the same slot to utilize the resources efficiently. Moreover, in the ISAC NR frame structure, REs that previously occupied by CSI-RS can now be replaced by actual downlink data. The frame structures of conventional NR and ISAC NR in one radio frame and one resource block are compared in Fig. \ref{connected}. Based on the frame structure, the throughput of 5G NR can be expressed as:
\begin{equation} \label{40}
\begin{aligned}
 \text{Throughput}\left(\text{in Mbps}\right) = 10^{-6}\cdot \sum_{j=1}^J\left(N_{\text{Layers}}^{\left(j\right)} \cdot Q_{\text{M}}^{\left(j\right)} \right.\\
 \left. \cdot \frac{N_{\text{PRB}}^{\text{BW}\left(j\right),\mu}\cdot12}{T_{\text{s}}^\mu} \cdot \left(1-\text{BER}^{\left(j\right)}-\text{OH}^{\left(j\right)}\right) \right)
\end{aligned}
\end{equation}
where $J$, $N_{\text{Layers}}$, $Q_{\text{M}}$, $N_{\text{PRB}}^{\text{BW},\mu}$, $T_{\text{s}}^\mu$ denote the number of carriers in carrier aggregation,  number of layers in MIMO, modulation order, number of practical resource blocks, and the average OFDM symbol duration, respectively. Moreover, $\text{BER}$ and $\text{OH}$ represent the bit error rate and the overhead percentage. 

\begin{figure*}[htb]
\centering
\subfloat[]{\includegraphics[width=3.5in]{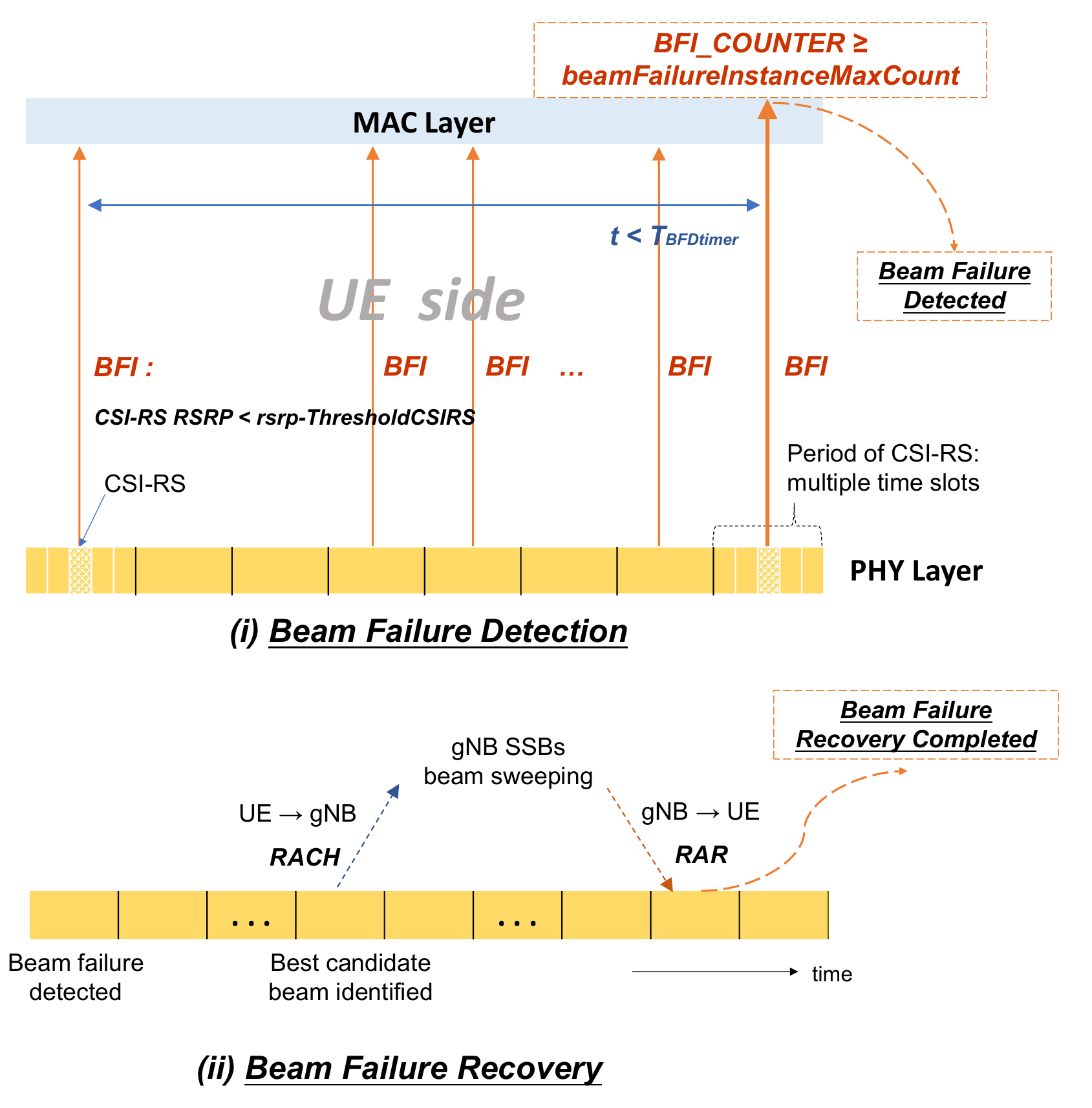}%
\label{bf_comm}}
\hfil
\subfloat[]{\includegraphics[width=3.5in]{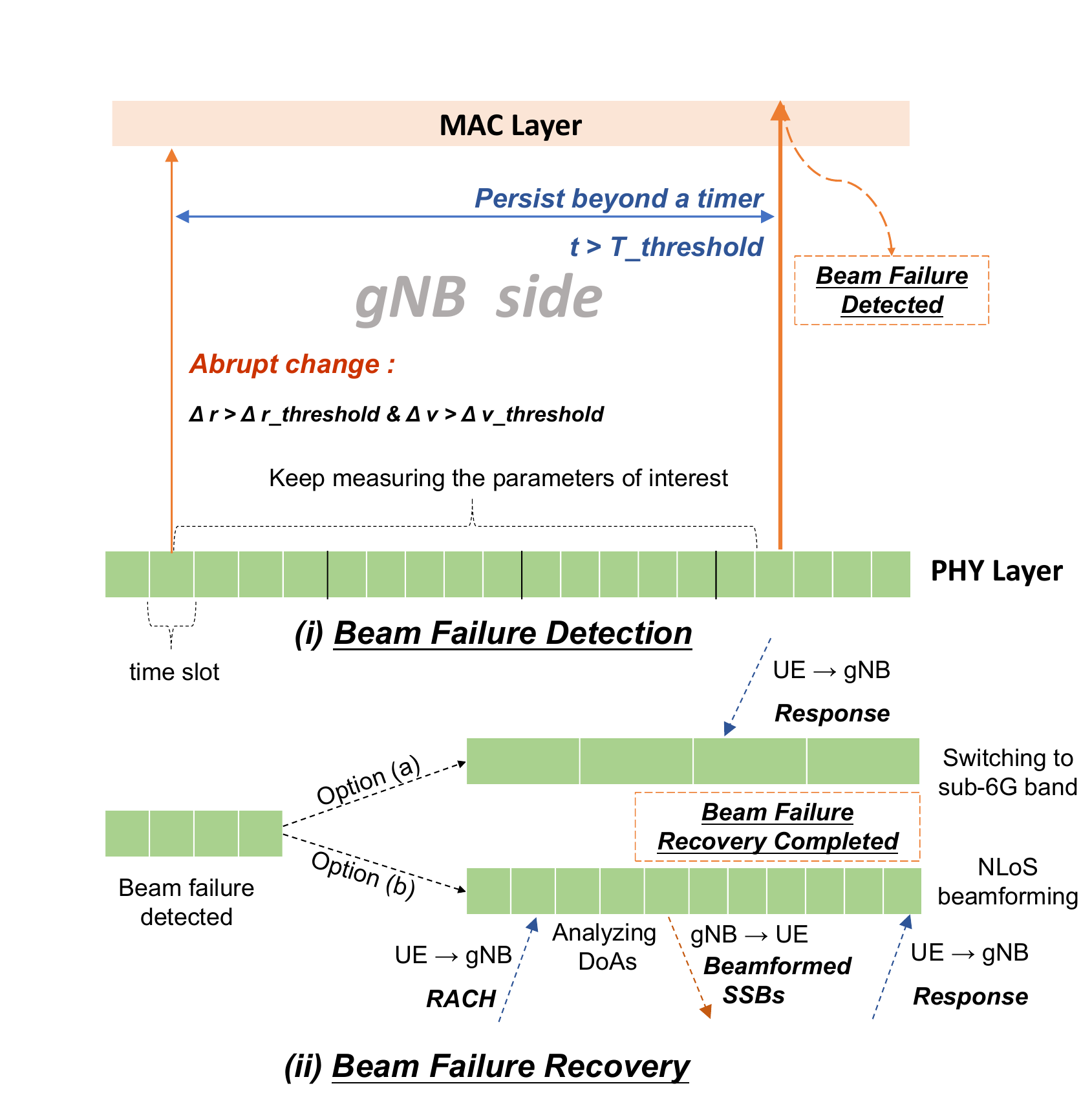}%
\label{bf_isac}}
\caption{Beam failure detection and recovery in conventional communication-only NR frame structure and proposed ISAC frame structure. (a) Communication-only. (b) ISAC.}
\label{bf}
\end{figure*}

Compared with the conventional communication-only NR frame structure, the ISAC V2I NR frame offers significant reductions in overhead caused by beam training and reference signals. Firstly, four dedicated time slots are allocated for beam training and synchronization purposes in conventional frame structure. However, this number is reduced to just one time slot with only synchronization purpose. Consequently, the overhead incurred by beam training is reduced up to $75\%$. Secondly, both DMRS and CSI-RS occupy a certain number of REs within a period and a RB in the conventional frame structure, specifically, 42 REs and 32 REs respectively. While in ISAC frame structure, the overhead caused by CSI-RS is reduced, leading to an overall overhead reduction of $32/(42+32)=43.24\%$. Furthermore, the tracking algorithm like EKF only utilizes the received echo for analysis, which eliminates the need for an uplink feedback in the ISAC V2I NR frame structure. By reducing the number of beam training slots, optimizing the deployment of reference signals and eliminating the need for an uplink feedback, the ISAC V2I NR frame structure improves resource utilization and enhances the efficiency of V2I communication in the connected mode.

\subsection{Beam Failure and Recovery} 

The radio link quality in NR is susceptible due to the high mobility of users, narrow beamwidth of pencil-like beams or the blockage between the gNB and the UE. These kinds of degradation of the radio link quality will lead to the occurrence of beam failure. In order to cope with this frequent occurring situation, beam failure recovery (BFR) procedures are introduced in NR to help identify new beam pairs and recover from poor communication quality. In this subsection, procedures of beam failure and recovery are compared between conventional communication-only scheme and proposed ISAC scheme, as depicted in Fig. \ref{bf}.

\subsubsection{BFR in conventional NR}
BFR in 5G NR takes two steps to complete, beam failure detection and beam failure recovery. 

\begin{itemize}[leftmargin=*]
    \item {\it Beam failure detection:} Beam failure detection is accomplished by the collaboration between physical layer (L1) and MAC layer (L2). UE monitors the radio link quality constantly by the measurement of L1 RSRP. Once the L1 RSRP of the current beam is below a certain threshold \textit{SS-rsrp} or \textit{CSIRS-rsrp}, L1 provides an indication of beam failure instance (BFI) to L2. L2 starts a timer as soon as it receives the first BFI and increases the BFI counter by $1$ for every BFI it receives. L2 will trigger the beam failure instantly if the BFI counter reaches to a certain threshold before the timer $T_{\text{BFDtimer}}$ expires \cite{3gpp.38.321}, i.e. \textit{BFI\_COUNTER} $\geq$ \textit{BFI\_{max}}.

    \item {\it Beam failure recovery:} After detecting beam failure, the recovery process is initiated by identifying candidate beams that surpass a specified threshold for recovery based on their L1 RSRP. UE performs random access channel (RACH) using the best beam from the previous selection and waits for the random access response (RAR) from the gNB to complete the beam failure recovery process. Note that the radio link failure might be declared if the recovery time exceeds or the communication quality of the new beam pair still suffers at a low level.
\end{itemize}

\subsubsection{BFR in ISAC NR}
The beam failure detection in conventional communication-only requires constant monitoring on the L1 RSRP of either SSB or CSI-RS, which are only being sent periodically. Consequently, the assessment of the radio link quality occurs only once within each period. This approach, however, incurs a substantial time cost in identifying beam failure due to the minimum period of CSI-RS and SSB being 4 time slots. 

To address this issue, we further propose to detect the beam failure based on monitoring the kinematic parameters of the target, in order to fully reap the active sensing capability. Specifically, the abrupt change in parameters such as range and velocity implies sudden blockage between the UE and the gNB in V2I network when the UE is being successfully tracked in previous time slots. Hence, when both parameters exhibit sudden variations exceeding certain thresholds, i.e. $\Delta r_ \text{th}$, $\Delta v_ \text{th}$ and persist beyond a specific number of time slots $T_ \text{th}$, beam failure between UE and gNB can be detected. 

The recurrent beam failure in mmWave band happens because the mmWave wave is susceptible to physical obstructions and suffers from high path loss. Here, we present two potential approaches for beam failure recovery process in mmWave V2I network. Firstly, in order to preserve the quality of communications, the gNB can adopt the strategy of switching to sub-6G band and transmitting omnidirectional signals, thereby mitigating the adverse effects of high attenuation caused by mmWave wave. Secondly, to maintain high data rates and throughput, gNB could leverage the channel reciprocity by analyzing the angle of arrivals from the uplink signal and beamform multiple data streams based on the DOAs of the NLoS paths.

\section{Link-Level Simulations}
The simulation scenario depicted in Fig. \ref{scenario} represents a vehicle driving along a straight road through real building clusters in the city of Shenzhen. The wireless communication channel between the vehicle and the gNB is modeled as a clustered delay line channel, comprising both LoS and NLoS paths. In the communication-only simulation case, the gNB is equipped with the UPA of size $8\times8$ and employs 32 CSI-RS antenna ports. The transmit array is divided into 4 subarrays in both the horizontal and vertical directions, and each subarray utilizes 4 oversample DFT beams in both directions which forms up the type-I CSI-RS codebook to enable beamforming. In terms of positioning, the gNB is located at the origin of the coordinate system, with the UPA's height of 8 meters. The vehicle's initial position is given as (25m, 40m, 1m). Additionally, the vehicle is driving at a constant speed of 20m/s with fluctuations. The simulation time of the whole process is 4 seconds, which consists of 32000 time slots. Other simulation parameters are specified in Table \ref{Sim}.

\begin{table}[!ht]
    \caption{Parameters of Simulation} \label{Sim} 
    \normalsize
    \centering
    \begin{tabular}{p{2cm} p{1.5cm} p{2cm} p{1.5cm}} 
    \hline
    \hline
    Parameter & Value & Parameter & Value\\
    \hline
    $f_c$ & $35\operatorname{GHz}$ & $T_{\text{max}}$ & $4\operatorname{s}$\\
    $\Delta f$ & $120\operatorname{kHz}$ & $\Delta T$ & $0.125\operatorname{ms}$\\
    $\sigma_\theta$ & $10^{-3}\operatorname{rad}$ & $\Tilde{\sigma}_\theta$ & $0.1\operatorname{rad}$\\
    $\sigma_d$ & $10^{-3}\operatorname{m}$ & $\Tilde{\sigma}_d$ & $0.2\operatorname{m}$\\
    $\sigma_v$ &$10^{-3}\operatorname{m/s}$ & $\Tilde{\sigma}_v$ & $0.15\operatorname{m/s}$\\
    $N_t, N_r$ & $64\;(8\times 8)$ & $M_r$ & $16\;(4\times 4)$\\
    $N_{\operatorname{Layers}}$ & $1$ & $Q_{\text{M}}$ & $4$\\
    $N_{\text{PRB}}^{\text{BW},\mu}$ & $208$ & $T_{\text{s}}^\mu$ & $8.929$ \textmu \text{s}\\
    $T_{\text{BFDtimer}}$ & $7.5\operatorname{ms}$ & \textit{BFI\_{max}} & $6$\\
    $\Delta v_{\text{th}}$ & $1\operatorname{m/s}$ & $\Delta r_{\text{th}}$ & $2\operatorname{m}$\\
    $T_{\text{th}}$ & $12 \operatorname{slots}$ & $P_{\text{FA}}$ & $0.01$\\
    \hline
    \end{tabular}
\end{table}

\subsection{Performance Comparison in Initial Access}
In this subsection, we present a comparative analysis of the initial access performance between the communication-only scheme and the ISAC scheme. In the simulation scenario, the vehicle commences the trajectory from randomly selected positions along the straight road at arbitrary points, within a time window during 0 to 20\text{ms}. In the communication-only scheme, beam training utilizing 64 SSB beams requires a time duration of 5ms out of a total 20ms period and identifies the optimal beam based on the SS-RSRP. By contrast, the ISAC scheme takes advantage of the inherent sensing capabilities of the system. It employs radar-based measurements to detect the presence and kinematic parameters of vehicles. An aggregation of 10 slots is employed for phase accumulation to ensure accurate kinematic parameter estimation. Subsequently, precise directional synchronization signals are transmitted to facilitate synchronization between the target vehicle and the gNB.

\begin{figure}[htbp]
\centering
\includegraphics[width=\columnwidth]{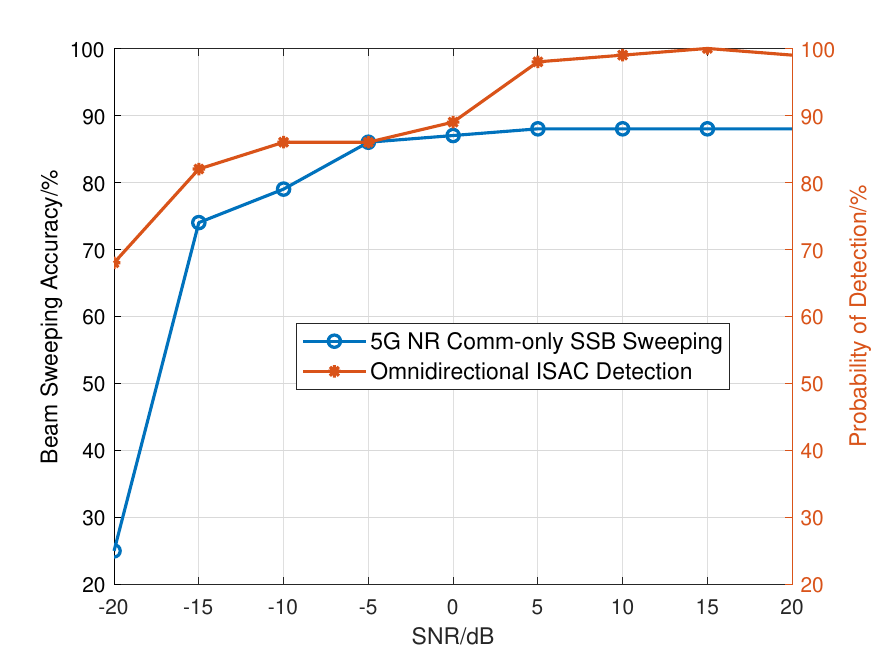}
\caption{Beam sweeping accuracy in conventional communication-only scheme and radar detection rate in ISAC scheme under different SNR.}
\label{IA_detection}
\end{figure}

\begin{figure}[htbp]
\centering
\includegraphics[width=\columnwidth]{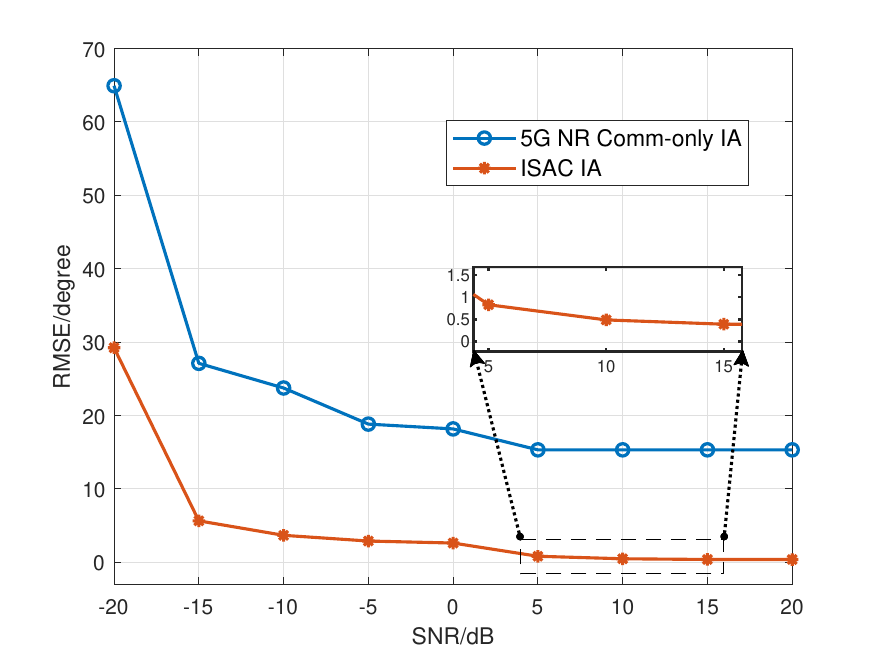}
\caption{RMSE of IA in communication-only scheme and ISAC scheme under different SNR.}
\label{IA_RMSE}
\end{figure}

To evaluate the effectiveness of ISAC scheme in IA, we first present the radar detection probability against the SSB sweeping accuracy in Fig. \ref{IA_detection}. In practical scenarios, the presence of noise introduces variability in the measurement of communication reference signal power and the radar detection of targets, leading to potential inaccuracies and missing detection. To gauge the precision of the beam angle, we employ the Root Mean Squared Error (RMSE) as a pertinent metric. Fig. \ref{IA_RMSE} shows the comparison between the RSRP-based and the radar-based beam identification. When the radar fails to detect the target's presence, the communication-only IA procedure is engaged. Consequently, the RMSE within the ISAC scheme can be considered as a weighted combination of two distinct scenarios: the RMSE under the condition of a missed radar detection, which corresponds to the communication-only RMSE, and the RMSE associated with a successful radar detection, reflecting the errors introduced by the radar system. Thus, the RMSE in the ISAC scheme represents a weighted aggregation of these two distinct situations. In contrast to traditional approaches that rely solely on predefined angles within the grid of SSB beams, the ISAC scheme leverages radar detection capabilities, significantly enhancing the system's proficiency in precisely ascertaining the location and dynamics of the vehicle during the initial access phase. This supplementary information plays a crucial role in optimizing the synchronization process between the target and the gNB, thereby improving the overall performance of the system.

\subsection{Performance Comparison in Connected Mode}

\begin{figure}[htbp]
\centering
\includegraphics[width=\columnwidth]{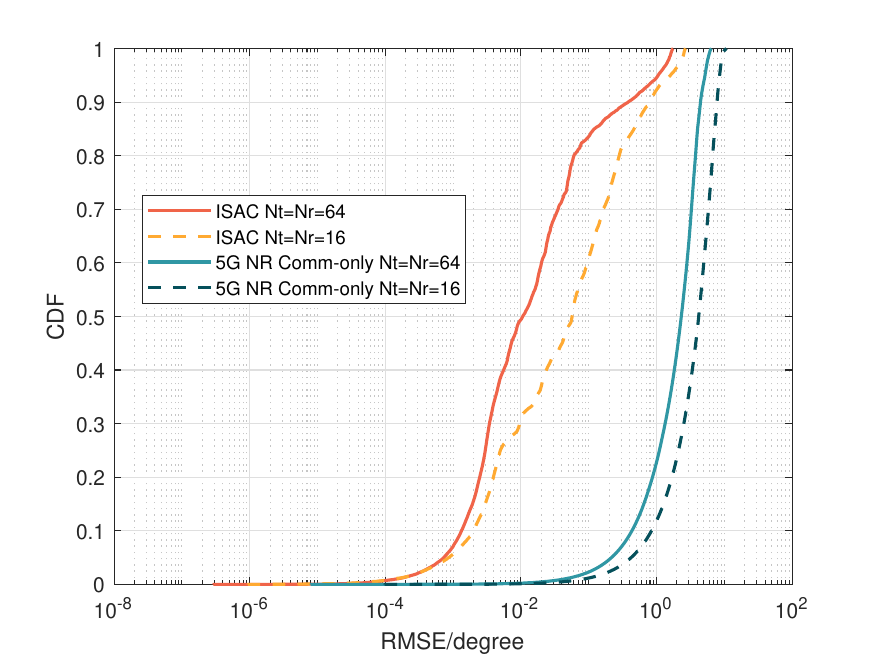}
\caption{Angle tracking performance of conventional communication-only scheme and ISAC scheme with initial state SNR=25dB.}
\label{angle_cdf}
\end{figure}

\begin{figure}[htbp]
\centering
\includegraphics[width=\columnwidth]{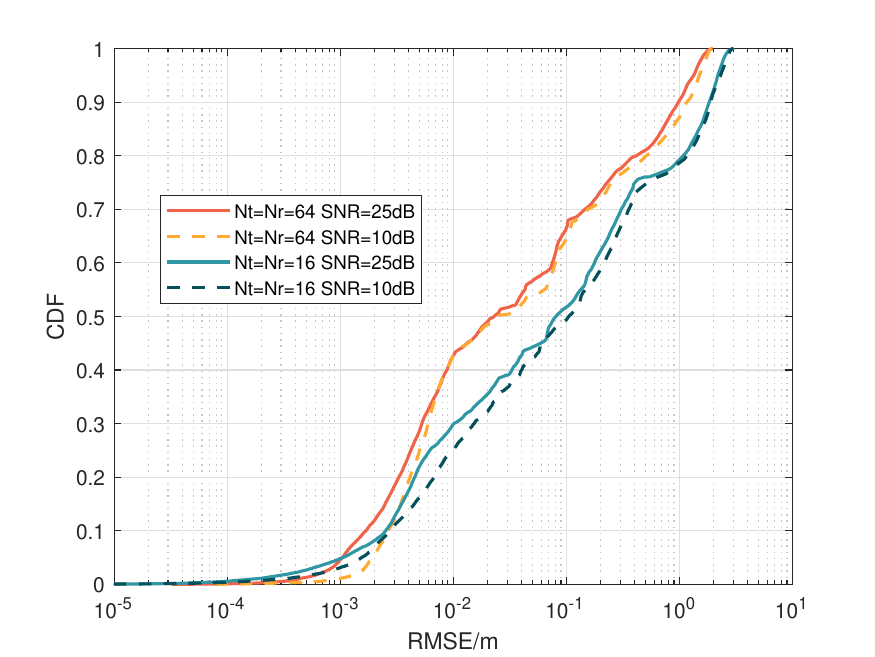}
\caption{Range tracking performance for ISAC scheme under different array size and different initial state SNR.}
\label{range_cdf}
\end{figure}

In this subsection, we present a comparative analysis of the performance between the conventional communication-only frame structure and the proposed ISAC frame structure in the connected mode. First, we evaluate the tracking performance of angle and range in Fig. \ref{angle_cdf} and Fig. \ref{range_cdf} in terms of the cumulative distribution function (CDF) of the RMSE. Angles in the communication-only scheme obey the standard codebook, which leads to large quantization errors. Through the prediction and estimation procedures in the EKF, precise tracking of angles is achieved, resulting in more accurate beamforming towards the target. The ISAC scheme offers advantages over the conventional communication-only scheme by enabling real-time measurement and prediction of the parameters of interest based on the information carried by the echoes of the signals. This approach enhances the flexibility and precision of tracking. Additionally, the range tracking exhibits similar performance to the angle tracking, with larger number of antennas achieves more precise tracking.

\begin{figure}[htbp]
\centering
\includegraphics[width=\columnwidth]{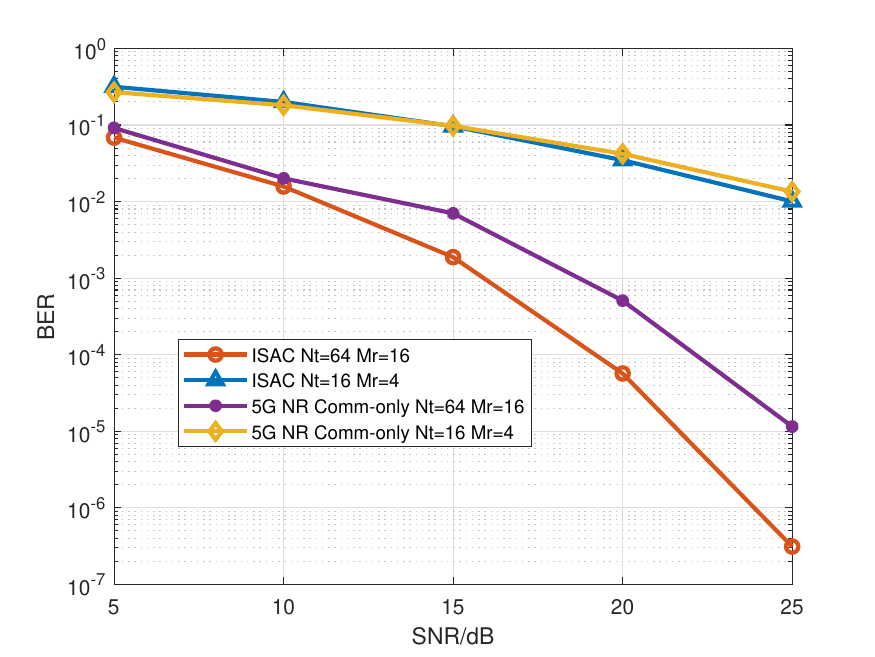}
\caption{BER comparison between conventional communication-only scheme and ISAC scheme in connected mode.}
\label{ber_cn}
\end{figure}

\begin{figure}[htbp]
\centering
\includegraphics[width=\columnwidth]{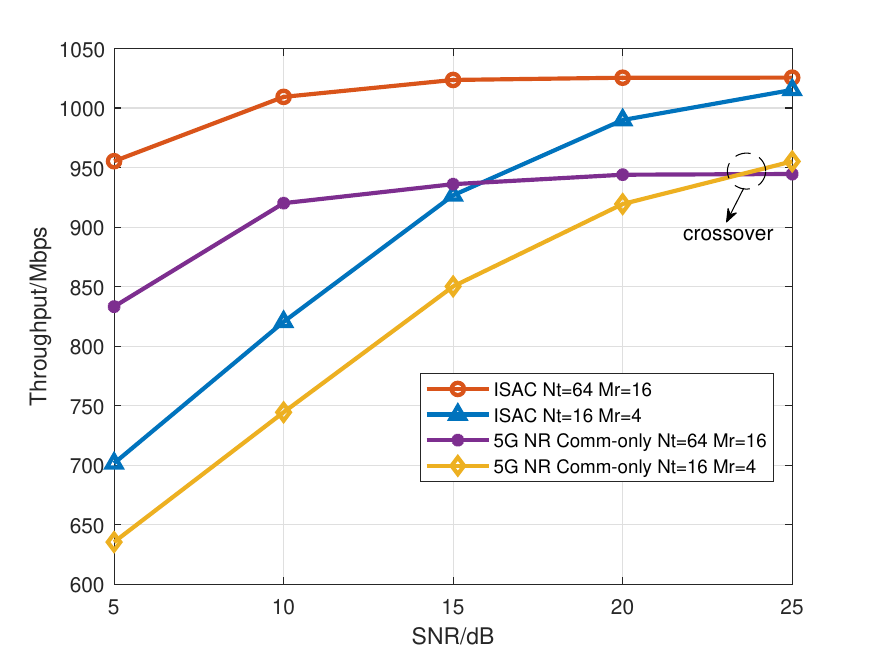}
\caption{Throughput comparison between conventional communication-only and ISAC scheme in connected mode.}
\label{tp_cn}
\end{figure}

In Fig. \ref{ber_cn} and Fig. \ref{tp_cn}, we present a comparison of the communication performance between the communication-only scheme and ISAC scheme using different metrics, i.e. BER and throughput. Fig. \ref{ber_cn} demonstrates a substantial improvement in BER achieved by the ISAC scheme compared to the communication-only scheme, particularly in the case of 64-antenna systems. Additionally, the presence of array gain leads to a significant overall improvement in BER performance in the 64-antenna scenario when compared to the 16-antenna case. Furthermore, we evaluate the performance of the throughput, which is defined by (\ref{40}). It is important to note that the throughput is influenced by several factors, such as the number of CSI-RS antenna ports and the percentage of beam training. At lower SNR, the 64-antenna communication-only configuration exhibits superior performance to the 16-antenna ISAC configuration, attributed to antenna gain, whereas with increasing SNR, the ISAC 16-antenna setup outperforms the communication-only configuration due to its reduced overhead. Overall, the ISAC scheme exhibits a significant improvement in throughput due to the reduced beam training overhead caused by SSBs, the elimination of reference signals like CSI-RS, and a slight improvement in BER. Notably, a crossover phenomenon is observed at high SNR in the communication-only 64-antenna and 16-antenna cases. This phenomenon arises because the number of CSI-RS antenna ports should be smaller than the actual number of antennas. As a result, only 16 CSI-RS antenna ports are used in the 16-antenna communication-only case, leading to reduced overhead caused by reference signals that slightly improves the throughput. Moreover, it is noteworthy that the ISAC scheme affords an additional reduction in overhead by eliminating the transmission of CSI-RS feedback in the uplink, a facet not explicitly depicted in the comparative analysis.

\subsection{Performance Comparison in Beam Failure and Recovery}

\begin{figure}[htbp]
\centering
\includegraphics[width=\columnwidth]{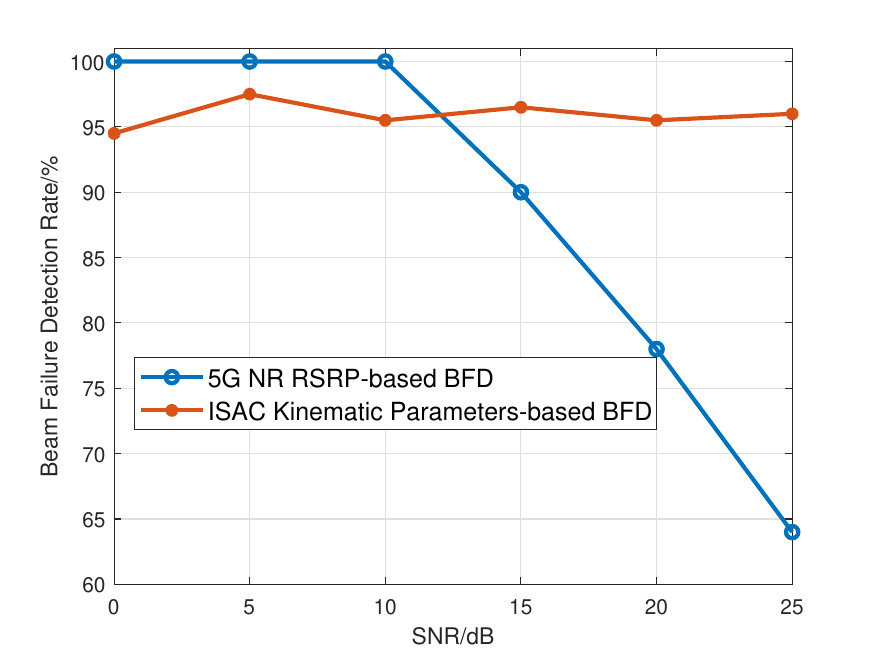}
\caption{Beam failure detection rate comparison between conventional communication-only scheme and ISAC scheme.}
\label{detection_possibility}
\end{figure}

\begin{figure}[htbp]
\centering
\includegraphics[width=\columnwidth]{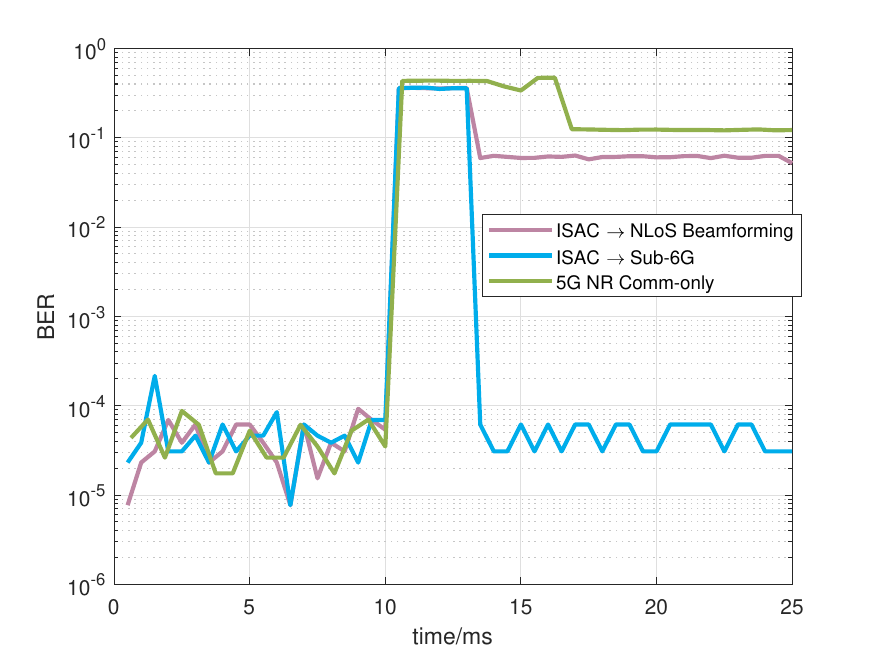}
\caption{BER comparison between conventional communication-only scheme and ISAC scheme in beam failure and recovery.}
\label{ber_bf}
\end{figure}

\begin{figure}[htbp]
\centering
\includegraphics[width=\columnwidth]{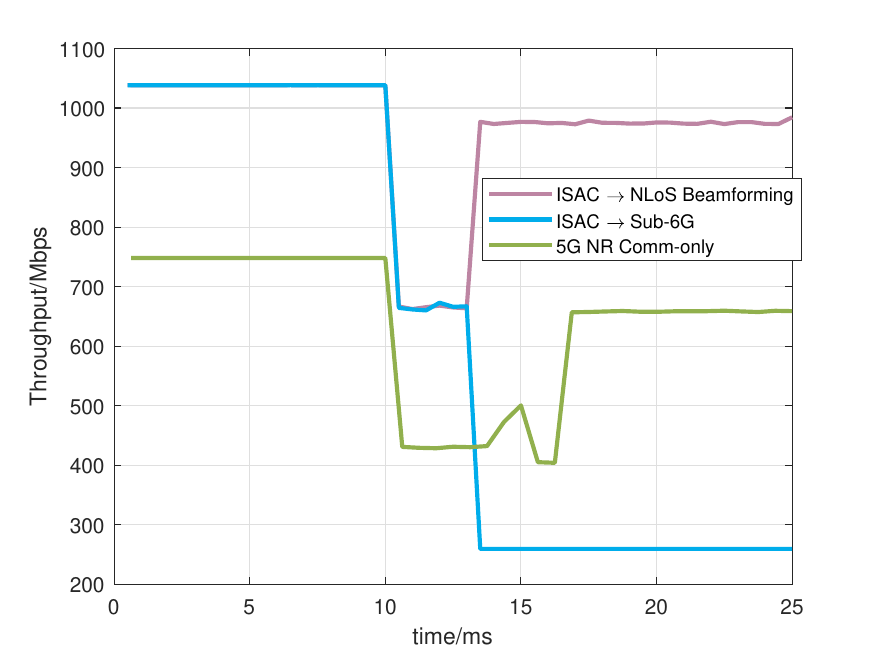}
\caption{Throughput comparison between conventional communication-only and ISAC scheme in beam failure and recovery.}
\label{tp_bf}
\end{figure}

Accurate and expeditious detection of failures and timely recovery procedures are crucial considering the high mobility of vehicles and the highly dynamic channel conditions in V2I networks. Taking into consideration the noise power spectral density of $-174$\text{dBm/Hz} and the bandwidth of $300$\text{MHz}, the noise power is calculated to be $-89.23$\text{dBm}. The RSRP threshold for identifying beam failure is set at a level $3$\text{dB} higher than the ambient noise floor. The beam failure scenario is designed to simulate a situation where a larger and faster vehicle obstructs the LoS path between the gNB and the tracking target, leading to communication between the gNB and the target relying exclusively on NLoS paths. The comparative analysis of beam failure detection probability under different SNR is presented in Fig. \ref{detection_possibility}, contrasting the communication-only RSRP-based approach and ISAC kinematic parameters-based approach. Notably, the RSRP-based scheme exhibits a diminishing trend in detection rates with increasing SNR. This trend can be attributed to the circumstance where, despite the LoS path being obstructed, the received signal from NLoS paths continues to surpass the established threshold, rendering beam failure detection challenging. Concurrently, communication quality remains compromised. Conversely, the kinematic parameters-based scheme demonstrates consistent performance across diverse SNR levels, maintaining its efficacy in failure detection.

Communication performance evaluation of beam failure and recovery is carried out for both the communication-only and ISAC approaches in Fig. \ref{ber_bf} and Fig. \ref{tp_bf}. The simulation of beam failure and recovery was conducted with the receive SNR at $20$\text{dB}.  Based on the 5G NR specifications, we assume that out of a total timer of $7.5$\text{ms},  $6$ or over BFIs can be identified as beam failures. Therefore, a minimum of $6$ periods of CSI-RS are required to detect the beam failure, which corresponds to a time duration of $3.75$\text{ms}. In the ISAC scheme, under the constant measure of the kinematic parameters of the target, we assume that an abrupt change in both velocity and range that persists for at least $12$ slots within a $20$-slot timer can be identified as beam failure. In Fig. \ref{ber_bf}, the results show that the communication-only scheme takes approximately $5$\text{ms} to identify the beam failure after it occurs, whereas the ISAC scheme achieves the same detection in just $2.5$\text{ms}. This significant reduction in detection time highlights the substantial improvement provided by the ISAC scheme. 

After the beam failure detection, the recovery process in communication-only scheme involves performing beam training to identify a new beam for re-establishing the connection. The time duration of the beam training process is not presented in Fig. \ref{ber_bf} or Fig. \ref{tp_bf} since it does not involve calculations of BER and throughput. In contrast, the proposed first solution in the ISAC scheme recovering by switching to sub-6G with the carrier frequency of $5$\text{GHz} and the numerology of $\mu=1$. This change in the frame structure leads to the drop-off in the throughput. However, due to the smaller path loss in sub-6G, the BER achieves better performance compared to mmWave. The ISAC scheme also presents a second solution that utilizes the analysis of the DOAs of possible NLoS paths. By employing beamforming in the angles of the NLoS paths, the BER and throughput are improved compared with the traditional beam training in communication-only scheme recovery process.

\section{Conclusion}
In this paper, we have proposed improved NR frame structures for both initial access and connected mode in ISAC-based NR V2I networks. By efficiently employing the sensing ability and tracking algorithm like EKF, the utilization of pilot and reference signals is minimized, resulting in improved performance in localization, tracking and communication. Moreover, we have introduced efficient schemes for rapid beam failure detection and recovery by monitoring abrupt changes in the kinematic parameters of targets. The algorithm aims to detect beam failures promptly and restore the connection within a short time duration, while striving to maintain high communication quality. Finally, to validate the effectiveness of the proposed frame structures and algorithms, numerical results of link-level simulations have been provided, showing that the resultant communication performance in BER and throughput outperforms the conventional NR frame structures and transmission protocols.

\bibliographystyle{IEEEtran}
\bibliography{IEEEabrv,ref}

\end{document}